\documentclass[aps,pra,twocolumn,groupedaddress,showpacs,showkeys]{revtex4-1}
\usepackage{amsmath,amssymb}
\usepackage{mathtools}
\usepackage{rotating}

\usepackage[usenames,dvipsnames,table]{xcolor}
\newcommand\y{\cellcolor{yellow!15}}
\newcommand\g{\cellcolor{green!15}}

\newcommand\f{\cellcolor{blue!15}}

\newcommand\sq{\square}
\newcommand\lt{\lhd}
\newcommand\rt{\rhd}

\begin{document}

%\preprint{}

\title{Three-Photon Stokes Mueller Polarimetry}


\author{Masood Samim, Serguei Krouglov and Virginijus Barzda}
\email[]{virgis.barzda@utoronto.ca}
%\thanks{}
\affiliation{Department of Chemical and Physical  Sciences, Department of Physics and Institute for Optical Sciences, University of Toronto. 3359 Mississauga Road North, Mississauga, Ontario L5L1C6, Canada}


\date{\today}

\begin{abstract}
The generalized theory of Stokes Mueller polarimetry is employed to develop the third-order optical polarimetry framework for third-harmonic generation (THG). The outgoing and incoming radiations are represented by 4-element and 16-element column vectors, respectively, and the intervening medium is represented by a $4\times 16$ triple Mueller matrix. Expressions for the THG Stokes vector and the Mueller matrix are provided in terms of coherency and correlation matrices, and expanded by four-dimensional $\gamma$ matrices that are analogues of Pauli matrices. Useful expressions of triple Mueller matrices are presented for cylindrically symmetric and isotropic structures. In addition, the relation between third-order susceptibilities and the measured triple Mueller matrix is provided. This theoretical framework can be applied for structural investigations of crystalline materials, including biological structures.
\end{abstract}


\maketitle

\section{Introduction}
Three-photon processes such as third-harmonic generation (THG) reveal unique structural information about the sample under study~\cite{Chu2004_Studi,Boyd2008_Nonli,Terhune1962_Optic,Tsang1995_Optic}. The generated third-harmonic signal from the material is related to the incoming radiation electric fields and determined by the third-order susceptibility tensor $\chi^{(3)}$ of the material~\cite{Boyd2008_Nonli}. The third-harmonic generation is an odd order process and has markedly different symmetry selection rules compared to the even order processes. 

In an optical setup the polarization-dependent interaction of light with matter can be described using Stokes Mueller, Poincar\'e or Jones formalism~\cite{Shurcliff1962_Polar,Kliger1990_Polar,Azzam1977_Ellip}. In the Stokes Mueller formalism, the light polarization is characterized by a Stokes vector, and its interaction with matter is represented by a Mueller matrix. The Stokes vector can describe partially- or completely-polarized light, and operates with intensities, which are real numbers, and thus, observables in an experiment.

The nonlinear Stokes Mueller equation for the third-order processes describes the relationship between the generated nonlinear signal radiation, the nonlinear properties of the media, and the incoming radiations:
%%%%% Eq: Stokes Poloarimetry  %%%%%
\begin{equation}\label{eq:NLStokeMueller}
s' (\omega_\sigma)= \mathcal{M}^{(3)}S^{(3)}(\omega_1,\omega_2,\omega_3)
\end{equation}
where $s'$ is the conventional $4\times 1$ Stokes vector of the generated radiation at $\omega_\sigma$ frequency and prime signifies the measured outgoing signal, while $S^{(3)}$ is a $16\times 1$ triple Stokes vector representing the polarization state of the three incoming electric fields that generate the light via nonlinear interactions. Henceforth, the $s'$ and $S^{(3)}$ are called the polarization state vectors for outgoing and incoming radiations, respectively. The triple Mueller matrix $\mathcal{M}^{(3)}$ describes the material properties of a third-order light-matter interaction. The Mueller matrix contains the nonlinear susceptibilities, which are independent of the incoming radiation intensities. The nonlinear Stokes Mueller polarimetry is applicable for non-ionizing radiations in the optical range (i.e. $I_{laser} << I_{atomic} = 4\times10^{20}W/m^2$), and the intensity independence of the Mueller matrix components can be tested by performing polarimetry measurements with several incoming radiation intensities~\cite{Samim2015_NLSM,Boyd2008_Nonli}.

%%%%% Fig: Three Photon Polarimetry %%%%%
\begin{figure}[htb]
	\includegraphics[width=0.48\textwidth]{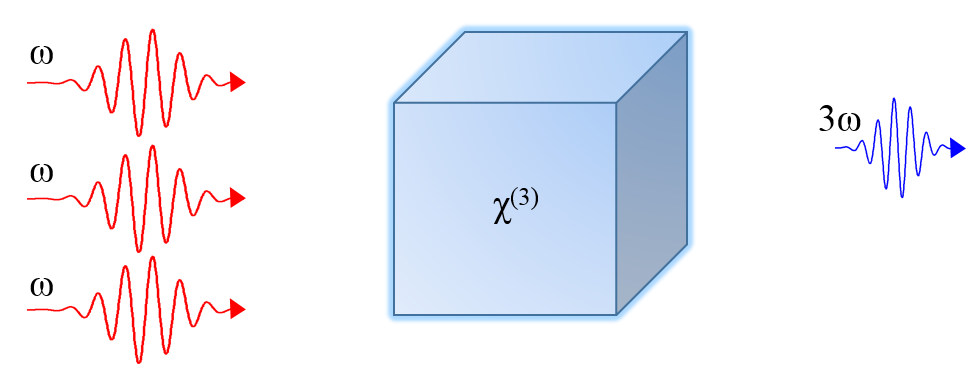}
	\caption[Schematic representation of a third-harmonic generation process analyzed by polarimetry.]{Schematic representation of a third-harmonic generation process analyzed by polarimetry. Three photons of incoming beam with frequency $\omega$ are incident onto the sample possessing the third-order susceptibility $\chi^{(3)}$, and the emitted radiation is at trice the incoming frequency, $3\omega$. In three-photon polarimetry, the incoming radiation is represented by the triple Stokes vector $S^{(3)}$, the medium is characterized with the matrix $\mathcal{M}^{(3)}$, and the outgoing measured signal is represented by a $4\times1$ Stokes vector $s'$.}\label{fig:ThreeoPhotonPolarimetry}
\end{figure}

In this paper, we concentrate on the three-photon processes, and develop specific equations for the THG from the generalized Stokes Mueller polarimetry formalism~\cite{Samim2015_NLSM}. The derivation will follow the same formalism that we used for the second harmonic generation process as well as the generalization for other nonlinear optical processes~\cite{Samim2015_NLSM,Samim2015_Doubl}. In Section~\ref{sec:Stokes} we will develop the equations for the outgoing and incoming radiation polarization states, followed by the expression for the triple Mueller matrix in Section~\ref{sec:NLMueller}. Specific examples of the Mueller matrix for real-valued susceptibilities with isotropic and hexagonal symmetries will be provided, which have relevance to the biological structures. The method for extracting susceptibility component values from the Mueller matrix will be provided in Section~\ref{sec:Chis}. The expressions for the triple Stokes Mueller polarimetric measurements will be presented in Section~\ref{sec:THGMeasurementPol}, including equations for the reduced polarimetry with linear polarization states of incoming and outgoing radiations.
\section{Derivation of Polarization State of Radiations}\label{sec:Stokes}
\subsection{Outgoing Radiation Stokes Vector}
The Stokes vector $s'$ for the outgoing third-harmonic radiation is characterized by a $4\times 1$ vector just as in the case for conventional Stokes vector. The coherency matrix of the third-harmonic signal is~\cite{Azzam1977_Ellip,Samim2015_NLSM}:
\begin{equation} 
\small C_{ab}'(3\omega)=\left<\Phi'(3\omega) \cdot \Phi'^{\dag}(3\omega)\right>_{ab}
	=\left<\Phi_a'(3\omega)\Phi_b'^{\dag}(3\omega)\right>
\end{equation} 
where $a$ and $b$ each run from 1 to 2, representing the orthogonal outgoing polarization orientations perpendicular to the light propagation direction, and $\Phi'(3\omega)$, which is directly proportional to the polarization density $P^{(3)}$, is the state (or simply the electric field) vector of the outgoing beam. The dagger symbol $\dagger$ denotes the complex conjugation and transposition. $\left< \cdot \right>$ signifies a time average over an interval long enough to make the time-averaging independent of the interval and fluctuations. Then, the outgoing radiation Stokes vector is~\cite{Samim2015_Doubl,Samim2015_NLSM,Azzam1977_Ellip}:
%%%%% Eq: Stokes Poloarimetry  %%%%%
\begin{equation}\label{eq:StokeNLOutgoing}
s'_t = {\rm{Tr}}\left( {{C}'{\tau_t^{}}} \right)= {{C}'_{ab}}{\left( {\tau_t^{}} \right)_{ba}} = \left<{\Phi'_a}\Phi_b'^*\right>{\left( {\tau_t^{}} \right)_{ba}} = \left<{\Phi '^\dag }{\tau_t }\Phi '\right>
\end{equation}
where $\tau_t$ ($t=0...3$) denotes the $2\times 2$ identity and Pauli matrices, which are hermitian and obey the unique orthogonality relation $\text{Tr}(\tau^{}_\mu \tau^{}_\nu)=2\delta^{}_{\mu\nu}$, where $\delta_{\mu\nu}$ is the Kronecker delta (see Appendix~\ref{sec:gamma}). 

The so-called degree of polarization ($dop$) is defined as~\cite{Born1999_Princ}:
\begin{equation}
	dop = \sqrt{{s'_1}^2+{s'_1}^2+{s'_1}^2}/s'_0
\end{equation}
The $dop$ ranges from 0 to 1, and can be used to measure the amount of scattered light in the radiation.
\subsection{Real-valued Triple Stokes Vector}
The third-order nonlinear signal is generated due to induced nonlinear polarization $ {P_i}^{(3)}$ in the material: 
\begin{equation}
P_i^{(3)} = \chi_{ijkl}^{(3)}{ E_j}{ E_k}{ E_l}= {\chi^{(3)}_{iA}}{\psi^{(3)}_A}
\end{equation}
where the first index $i$ indicates the direction of outgoing polarization, while the indices $j$, $k$ and $l$ indicate the direction of polarization of incoming electric fields, and summation is assumed over the repeated indices~\cite{Boyd2008_Nonli,Samim2015_NLSM}. The index $A$ is composed of the indices $j, k, l$ and depends on the third-order process being studied (i.e. THG, coherent anti-Stokes Raman scattering (CARS), etc.).

For triple Stoke Mueller polarimetry of THG, the incoming radiation electric fields state vector $\psi^{(3)}(\omega_1,\omega_2,\omega_3)$ has the same frequency $\omega$ for all three electric fields. Therefore the polarization state is:
\begin{equation}
\begin{array}{l}
\psi^{(3)}{(\omega,\omega,\omega)}  = \left( {\begin{array}{*{20}{c}}
	{{E_1}^3}\\
	{{E_2}^3}\\
	{3{\mkern 1mu} {E_1}^2{\mkern 1mu} {E_2}}\\
	{3{\mkern 1mu} {E_1}{\mkern 1mu} {E_2}^2}
	\end{array}} \right)
\end{array} 
\end{equation}
where the subscripts 1 and 2 represent the orthogonal electric field vector orientations forming the plane perpendicular to the light propagation. Thus, the coherency matrix for the three-photon process is:
\onecolumngrid
\begin{equation}\label{eq:TripleCoherency}
\begin{split}
&\rho^{(3)}{(\omega,\omega,\omega)}=\left<\psi^{(3)}\cdot \psi^{(3)\dag}\right>=\\
&\small
\left(\begin{array}{cccc} 
\left<{E_{1}}^3\, {{E_{1}^*}}^3\right> & \left<{E_{1}}^3\, {{E_{2}^*}}^3\right> & \left<3\, {E_{1}}^3\, {{E_{1}^*}}^2\, {E_{2}^*}\right> & \left<3\, {E_{1}}^3\, {E_{1}^*}\, {{E_{2}^*}}^2\right>\\ 
\left<{E_{2}}^3\, {{E_{1}^*}}^3\right> & \left<{E_{2}}^3\, {{E_{2}^*}}^3\right> & \left<3\, {E_{2}}^3\, {{E_{1}^*}}^2\, {E_{2}^*}\right> & \left<3\, {E_{2}}^3\, {E_{1}^*}\, {{E_{2}^*}}^2\right>\\ 
\left<3\, {E_{1}}^2\, E_{2}\, {{E_{1}^*}}^3\right> & \left<3\, {E_{1}}^2\, E_{2}\, {{E_{2}^*}}^3\right> & \left<9\, {E_{1}}^2\, E_{2}\, {{E_{1}^*}}^2\, {E_{2}^*}\right> & \left<9\, {E_{1}}^2\, E_{2}\, {E_{1}^*}\, {{E_{2}^*}}^2\right>\\ 
\left<3\, E_{1}\, {E_{2}}^2\, {{E_{1}^*}}^3\right> & \left<3\, E_{1}\, {E_{2}}^2\, {{E_{2}^*}}^3\right> & \left<9\, E_{1}\, {E_{2}}^2\, {{E_{1}^*}}^2\, {E_{2}^*}\right> & \left<9\, E_{1}\, {E_{2}}^2\, {E_{1}^*}\, {{E_{2}^*}}^2\right> \end{array}\right)
= \left(\begin{array}{cccc} C_{11}^3 & C_{12}^3 & 3\, C_{11}^2\, C_{12} & 3\, C_{11}\, C_{12}^2\\ C_{21}^3 & C_{22}^3 & 3\, C_{21}^2\, C_{22} & 3\, C_{21}\, C_{22}^2\\ 
3\, C_{11}^2\, C_{21} & 3\, C_{12}^2\, C_{22} & 9\, C_{11}^2\, C_{22} & 9\, C_{12}^2\, C_{21}\\ 
3\, C_{11}\, C_{21}^2 & 3\, C_{12}\, C_{22}^2 & 9\, C_{12}\, C_{21}^2 & 9\, C_{11}\, C_{22}^2 \end{array}\right)
\end{split}
\end{equation}
\twocolumngrid
where $C$ is the ordinary coherency matrix for the fundamental wavelength $\omega$~\cite{Born1999_Princ}. In Eq.~\ref{eq:TripleCoherency}, going from the first to the second matrix, we assume $\left<E_jE_kE_lE_m^*E_n^*E_o^*\right>=\left<E_jE_m^*\right>\left<E_kE_n^*\right>\left<E_lE_o^*\right>$ (each indices $j,k,l,m, n$ and $o$ run from 1 to 2), which allows to use the conventional coherency matrix to redefine the interacting electric fields. This assumption is justified if the incident radiation is coherent i.e from a laser source.

The general triple Stokes vector can be found similar to the linear and two-photon processes according to:
\begin{equation}\label{eq:StokeNLIncoming}
{S^{(3)}_N}(3\omega) = {\rm{Tr}}\left( {\rho^{(3)}}\, {\gamma_{_N}^{}} \right) 
	= \left<{\psi^{(3)}_A}\psi^{(3)*}_B \right> {\left( {\gamma_{_N}^{}} \right)_{BA}}
\end{equation}
where $A$ and $B = 1, \cdots, 4$, $N= 1,\cdots, 16$, and the $4 \times 4$ $\gamma$ matrices and the associated identity matrix are analogous to Pauli matrices. They are hermitian and obey the unique orthogonality relation $\text{Tr}(\gamma^{}_M \gamma^{}_N)=2\delta^{}_{MN}$ (see Appendix~\ref{sec:gamma}). Therefore, the elements of the triple Stokes vector describing the polarization state for incoming radiation are:
\onecolumngrid
\begin{equation}\label{eq:TripleStokes}
\small
\begin{split}
&{S^{(3)}_N(\omega,\omega,\omega)} =\\
&\left(\begin{array}{c} 
\frac{\sqrt{2}}{2} \left(\left<{E_{1}}^3\, {E_{1}^*}^3\right> + \left<9 {E_{1}}^2\, E_{2}\, {E_{1}^*}^2\, E_{2}^* \right>+ \left<9 E_{1}\, {E_{2}}^2\, E_{1}^*\, {E_{2}^*}^2\right> + \left<{E_{2}}^3\, {E_{2}^*}^3\right> \right)\\ 
\frac{\sqrt{6}}{6} \left( \left<{E_{1}}^3\, {E_{1}^*}^3\right>  + \left<{9  {E_{1}}^2\, E_{2}\, {E_{1}^*}^2\, E_{2}^*}\right>  - \left<{27 E_{1}\, {E_{2}}^2\, E_{1}^*\, {E_{2}^*}^2}\right>  + \left<{ {E_{2}}^3\, {E_{2}^*}^3}\right>  \right)\\ 
\frac{\sqrt{3}}{3} \left( \left<{E_{1}}^3\, {E_{1}^*}^3\right>  - \left<18 {E_{1}}^2\, E_{2}\, {E_{1}^*}^2\, E_{2}^*\right>  +  \left<{E_{2}}^3\, {E_{2}^*}^3\right>  \right)\\ 
\left<{E_{1}}^3\, {E_{1}^*}^3\right>  - \left<{E_{2}}^3\, {E_{2}^*}^3\right> \\ 
\left<{E_{1}}^3\, {E_{2}^*}^3\right>  + \left<{E_{2}}^3\, {E_{1}^*}^3\right> \\ 
\left<3\, {E_{1}}^2\, E_{2}\, {E_{2}^*}^3\right>  + \left<3\, {E_{2}}^3\, {E_{1}^*}^2\, E_{2}^*\right> \\ 
\left<9\, {E_{1}}^2\, E_{2}\, E_{1}^*\, {E_{2}^*}^2\right>  + \left<9\, E_{1}\, {E_{2}}^2\, {E_{1}^*}^2\, E_{2}^*\right> \\ 
\left<3\, E_{1}^*\, {E_{2}}^3\, {E_{2}^*}^2\right>  + \left<3\, E_{1}\, {E_{2}}^2\, {E_{2}^*}^3\right> \\ 
\left<3\, E_{2}^*\, {E_{1}}^3\, {E_{1}^*}^2\right>  + \left<3\, E_{2}\, {E_{1}}^2\, {E_{1}^*}^3\right> \\ 
\left<3\, {E_{1}}^3\, E_{1}^*\, {E_{2}^*}^2\right>  + \left<3\, E_{1}\, {E_{2}}^2\, {E_{1}^*}^3\right> \\ 
\left( \left<{E_{1}}^3\, {E_{2}^*}^3\right> - \left<{E_{2}}^3\, {E_{1}^*}^3\right> \right) \mathrm{i} \\ 
\left(\left<3\, {E_{2}}^3\, {E_{1}^*}^2\, E_{2}^*\right> - \left<3\, {E_{1}}^2\, E_{2}\, {E_{2}^*}^3\right> \right) \mathrm{i} \\ 
\left(\left<9\, {E_{1}}^2\, E_{2}\, E_{1}^*\, {E_{2}^*}^2\right> - \left<9\, E_{1}\, {E_{2}}^2\, {E_{1}^*}^2\, E_{2}^*\right> \right)\mathrm{i} \\ 
\left(\left<3\, {E_{2}}^3\, E_{1}^*\, {E_{2}^*}^2\right> - \left<3\, E_{1}\, {E_{2}}^2\, {E_{2}^*}^3\right> \right)\mathrm{i}  \\ 
\left(\left<3\, {E_{1}}^3\, {E_{1}^*}^2\, E_{2}^*\right> - \left<3\, {E_{1}}^2\, E_{2}\, {E_{1}^*}^3\right> \right)\mathrm{i}  \\ 
\left(\left<3\, {E_{1}}^3\, E_{1}^*\, {E_{2}^*}^2\right> - \left<3\, E_{1}\, {E_{2}}^2\, {E_{1}^*}^3\right> \right)\mathrm{i}  \end{array}\right)
=\frac{1}{4}\left(\begin{array}{c} 
\sqrt{2}s_{0} (5 {s_{0}}^2 - 3 {s_{1}}^2)\\  
\sqrt{6} (- \frac{4}{3}{s_{0}}^3 + 3\,  {s_{0}}^2\, s_{1} + 2\,  s_{0}\, {s_{1}}^2 - 3\,  {s_{1}}^3)\\  
\sqrt{3} (- \frac{8}{3}{s_{0}}^3 - 3\,  {s_{0}}^2\, s_{1} + 4\,  s_{0}\, {s_{1}}^2 + 3\,  {s_{1}}^3)\\ 
s_{1} (3\, {s_{0}}^2 + {s_{1}}^2)\\ 
s_{2} ({s_{2}}^2 - 3\, {s_{3}}^2)\\ 
3 (s_{0} - s_{1})({s_{2}}^2 - {s_{3}}^2)\\ 
9 s_{2} ({s_{2}}^2 + {s_{3}}^2)\\ 
3\, s_{2}\, {\left(s_{0} - s_{1}\right)}^2\\ 
3\, s_{2}\, {\left(s_{0} + s_{1}\right)}^2\\ 
3(s_{0} + s_{1})({s_{2}}^2 - {s_{3}}^2)\\ 
s_{3}(3\, {s_{2}}^2 - {s_{3}}^2)\\ 
- 6 s_{2}\, s_{3}\, \left(s_{0} - s_{1}\right)\\ 
9 s_{3} ({s_{2}}^2 + {s_{3}}^2)\\ 
- 3 s_{3}\, {\left(s_{0} - s_{1}\right)}^2\\ 
3 s_{3}\, {\left(s_{0} + s_{1}\right)}^2\\ 
6 s_{2}\, s_{3}\, \left(s_{0} + s_{1}\right) \end{array}\right)
\end{split}
\end{equation}
\twocolumngrid
Note, the last six elements of the vector above vanish when $s_3$ is zero. It implies that the triple Stokes vector for THG will have at most the first 10 nonzero elements if the incoming radiation is linearly polarized.

Similar to the linear Stokes parameters, the vector for third-order interaction obeys the following relation:	
\begin{equation}\label{eq:}
3S_1^2 \ge \sum\limits_{N = 2}^{16} {S_N^2} 
\end{equation}	
where the equality is valid for a purely polarized state. Therefore, it is convenient to use the degree of polarization (DOP) parameter to characterize the fundamental radiation:
\begin{equation}\label{eq:DOP3}
DOP(\omega,\omega,\omega) = \sqrt {\sum\limits_{N = 2}^{16} {S_N^2} /3S_1^2}
\end{equation}
where $DOP$ ranges from 0 to 1 for unpolarized to fully polarized fundamental radiation, respectively.
\section{Real-valued Triple Mueller Matrix $\mathcal{M}^{(3)}$ for Intervening Medium}\label{sec:NLMueller}
By substituting linear and nonlinear Stokes vector expressions (Eq.~\ref{eq:StokeNLOutgoing} and~\ref{eq:StokeNLIncoming}, respectively) into the general nonlinear polarimetry Eq.~\ref{eq:NLStokeMueller} the following expression is obtained:
\begin{equation}\label{eq:OutnInStates}
\left< {\Phi '^\dag }\,{\tau^{}_t }\,\Phi ' \right>= \mathcal{M}_{t N}^{(3)}\,\left< {\psi ^\dag }\,{\gamma_{_N}^{}}\,\psi\right>
\end{equation}
In this framework, each component of the vector $\Phi'$ of the generated electric field is proportional to the polarization state of incoming electric field and to the susceptibility tensor components of the material. By substituting explicit expressions of $\Phi'$ and $\Phi'^\dag$ into Eq.~\ref{eq:OutnInStates} in the elemental form the following equation is obtained:
%%%%% Eq: Stokes Poloarimetry  %%%%%
\begin{equation}\label{eq:NLStokeMuellerElments}
\left<  \chi^{(3)*}_{aA}\psi _A^*{\left( {{\tau^{}_t }} \right)_{ab}}{\chi^{(3)}_{bB}}{\psi _B} \right> = {\mathcal{M}^{(3)}_{t N}} \left< {\psi_A^*}\,({\gamma_{_N}})^{}_{AB}\,\psi^{}_B\right>
\end{equation}
where the contracted notation $A$ and $B = 1, \cdots, 4$ in $\chi^{(3)}_{iA}=\chi^{(3)}_{ijkl}$ is defined as:
\begin{equation}\label{eq:ContractedNotationNumerical}
\begin{array}{ccccc}
	jkl:&111&222&112,121,211&122,212,221\\
	A:&1&2&3&4\\
\end{array}
\end{equation}
and the same contracted notation is used for $B$. Note that here the matrix is constructed mainly with the THG process in mind, where the Kleinman symmetry is not required. In extending the matrix for other third-order processes it is required to ensure that the Kleinman symmetry is valid. Otherwise, an effective $\chi^{(3)}$ is defined, or the dimensions of the $\gamma$ matrices must be increased.

In a highly scattering media such as biological tissue, the system may not be completely coherent, and the source of the signal may be an ensemble of scatterers. Therefore, an ensemble average of individual elements with probability $p_e$ may be more appropriate to consider~\cite{Kim1987_Relat}. In addition, since Eq.~\ref{eq:NLStokeMuellerElments} is written in terms of individual elements, the state functions of the fundamental radiation can be dropped and the nonlinear Mueller matrix elements $\mathcal{M}_{tN}$ can be written in terms of the third-order susceptibilities as:
%%%%% Eq: Stokes Poloarimetry  %%%%%
\begin{equation}\label{eq:MuellerNLEnsemble2}
{\mathcal{M}^{(3)}_{t N}} =  
{\textstyle{1 \over 2}} \left<\chi^{(3)*}_{aA}{\chi^{(3)}_{bB}}\right>_e {\left( {{\tau^{}_t }} \right)_{ab}}{\left( {{\gamma_{_N}}} \right)_{BA}}
\end{equation}
where the ensemble average of the susceptibility values is expressed as:
$<\chi^{(3)*}_{aA}{\chi^{(3)}_{bB}}>_e = \sum_{e}^{} p_e 
(\chi^{(3)*}_{aA}{\chi^{(3)}_{bB}})$.
In deriving the Eq.~\ref{eq:MuellerNLEnsemble2} for the ensemble of $\chi^{(3)}$ the order of variables is a non-issue because both equations (Eqs.~\ref{eq:NLStokeMuellerElments} and~\ref{eq:MuellerNLEnsemble2}) are expressed in the elemental form. The correlation matrix $<\chi^{(3)*}_{aA}{\chi^{(3)}_{bB}}>_e$ contains information about the ensemble, and in the case of a homogeneous medium reduces to a product of $\chi^{(3)*}_{aA}{\chi^{(3)}_{bB}}$ of a single source.

Note that the outgoing radiation may not be fully polarized, since the generated light is no longer originating from a single source, but rather from an ensemble of sources. Thus, Eq.~\ref{eq:MuellerNLEnsemble2} is a better representation of experimental data from a heterogeneous medium.

In comparison to the linear Mueller matrix elements, which are composed of products of linear susceptibilities and Pauli matrices, the triple $\mathcal{M}^{(3)}$ is composed of nonlinear susceptibilities and $4\times 4$ $\gamma$ matrices. All elements of the matrix for the nonlinear interaction are real, a fact that leads to a very useful and a much-desired expression for determining the nonlinear susceptibilities, which will be shown in Section~\ref{sec:Chis}.

Similar to the symbolic notation that was given for the second-order matrix $\mathcal{M}^{(2)}$ (see ref.~\cite{Samim2015_Doubl}), the symbolic matrix for the third-order $\mathcal{M}^{(3)}$ becomes:
\onecolumngrid
%%%% Eq: Stokes Poloarimetry  %%%%%
\begin{equation}\label{eq:TripleMuellerSymbolic}
\arraycolsep=2pt
\left( {\begin{array}{cccccccccccccccc}
	{NP}&{NP}&{NP}&{NP}&\y{I_\diamondsuit^c}&\y{I_\Delta^c}&\y{I_\nabla^c}&\y{I_\lt^c}&\y{I_\rt^c}&\y{I_\sq^c}&\y{I_\diamondsuit^s}&\y{I_\Delta^s}&\y{I_\nabla^s}&\y{I_\lt^s}&\y{I_\rt^s}&\y{I_\sq^s}\\
	{NP}&{NP}&{NP}&{NP}&\y{I_\diamondsuit^c}&\y{I_\Delta^c}&\y{I_\nabla^c}&\y{I_\lt^c}&\y{I_\rt^c}&\y{I_\sq^c}&\y{I_\diamondsuit^s}&\y{I_\Delta^s}&\y{I_\nabla^s}&\y{I_\lt^s}&\y{I_\rt^s}&\y{I_\sq^s}\\
	\f{{O^c}}&\f{{O^c}}&\f{{O^c}}&\f{{O^c}}&\g{OI_\diamondsuit^c}&\g{OI_\Delta^c}&\g{OI_\nabla^c}&\g{OI_\lt^c}&\g{OI_\rt^c}&\g{OI_\sq^c}&\g{OI_\diamondsuit^s}&\g{OI_\Delta^s}&\g{OI_\nabla^s}&\g{OI_\lt^s}&\g{OI_\rt^s}&\g{OI_\sq^s}\\
	\f{{O^s}}&\f{{O^s}}&\f{{O^s}}&\f{{O^s}}&\g{OI_\diamondsuit^s}&\g{OI_\Delta^s}&\g{OI_\nabla^s}&\g{OI_\lt^s}&\g{OI_\rt^s}&\g{OI_\sq^s}&\g{OI_\diamondsuit^c}&\g{OI_\Delta^c}&\g{OI_\nabla^c}&\g{OI_\lt^c}&\g{OI_\rt^c}&\g{OI_\sq^c}\end{array}} \right)
\end{equation}
\twocolumngrid
The subscripts $s$ and $c$ denote the sin and cos of the relative phase between the susceptibility tensor elements that make up the triple Mueller matrix element. Phase-independent ($NP$) elements are present in the first two rows and four columns and are composed of the squares of susceptibility components; those dependent on the incoming $I$ index are situated in the first two rows and last twelve columns; elements dependent on the outgoing $O$ index are located in the last two rows and first four columns; and the elements that depend on incoming as well as outgoing $OI$ indices form the remaining elements.

Additionally, the following symmetry features of the triple Mueller matrix can be seen: Columns five to ten of the triple Mueller matrix ($N=5,\cdots,10$) have the same susceptibility products as columns eleven to sixteen ($N+6$), respectively, where the retardedness of the components in the last six columns acquire additional $\pi/2$ phase shift relative to the six columns before them (superscripts $c$ and $s$). The corresponding columns with the same susceptibilities are indicated as subscripts $\diamondsuit$, $\Delta$, $\nabla$, $\lt$, $\rt$ and $\square$ in the symbolic matrix~\ref{eq:TripleMuellerSymbolic}. In the following, explicit expressions of the triple Mueller matrix for a few special cases will be presented.
\subsection*{Triple Mueller Matrix for Real Susceptibilities}
When susceptibilities are real, the Mueller matrix components that have sin dependency on the retardance phase reduce to zero (see the components with superscript $s$ in the symbolic Mueller matrix~\ref{eq:TripleMuellerSymbolic}). Therefore, the last six columns of the first three rows and the first ten elements of the last row in $\mathcal{M}^{(3)}$ are zero. In addition, if Kleinman symmetry is valid for a third-order process, then there are five unique nonzero susceptibilities. The Mueller matrix can be expressed by using four unique susceptibility component ratios as follows:
\onecolumngrid
\setcounter{MaxMatrixCols}{20}
%%%% Eq: Tripel Stokes Poloarimetry  %%%%%
\begin{equation}\label{eq:TripleMuellerRatio}
\small\arraycolsep=1pt % default: 5pt
\medmuskip = 1mu % default: 4mu plus 2mu minus 4mu
\begin{split}
&{\mathcal{M}_{THG,Kleinman}^{(3)} ={(\chi^{(3)}_{14})}^2}\\
&\left(\begin{matrix}
\frac{\sqrt{2}}{4} \left(a^2 + b^2 + 2\, c^2 + 2\, d^2 + 2\right) & \frac{\sqrt{6}}{12} \left(a^2 + b^2 + 2\, c^2 - 2\, d^2 - 2\right) & \frac{\sqrt{3}}{6} \left(a^2 + b^2 - c^2 + d^2 - 2\right)& ({a^2} - {b^2} + {c^2}- {d^2})/2 \\
\frac{\sqrt{2}}{4} \left(a^2 - b^2\right) & \frac{\sqrt{6}}{12} \left(a^2 - b^2 + 4\, d^2 - 4\right) & \frac{\sqrt{3}}{6} \left(a^2 - b^2 - 3\, c^2 + d^2 + 2\right) &  ({a^2}+{b^2} - {c^2} - {d^2})/2 \\
\frac{\sqrt{2}}{2} \left(c + d + a\, c + b\, d\right) & \frac{\sqrt{6}}{6} \left(c - 3\, d + a\, c + b\, d\right) & \frac{\sqrt{3}}{3} \left(a\, c - 2\, c + b\, d\right) & a\, c - b\, d\\
0 & 0 & 0 & 0 \\
\end{matrix}\right.\\
&\qquad \left.\arraycolsep=4pt
\begin{matrix}
a\, d + b\, c & c\, d + b & c + d & d\, \left(1 + b \right) & c\, \left(a + 1\right) & a + c\, d   & 0 & 0 & 0 & 0 & 0 & 0\\
a\, d - b\, c & c\, d - b & c - d & d\, \left(1 - b\right) &  c\, \left(a - 1\right) & a - c\, d & 0 & 0 & 0 & 0 & 0 & 0\\
a\, b + c\, d & d + b\, c & c\, d + 1 & d^2 + b & c^2 + a & c + a\, d & 0 & 0 & 0 & 0 & 0 & 0\\ 
0 & 0 & 0 & 0 & 0 & 0 & a\, b - c\, d  & d- b\, c & c\, d - 1 &  d^2 -b & a- c^2  & a\, d -c \\
\end{matrix}
\right)
\end{split}
\end{equation}
\twocolumngrid
where $a = \chi^{(3)}_{11}/\chi^{(3)}_{14}$, $b = \chi^{(3)}_{22}/\chi^{(3)}_{14}$, $c = \chi^{(3)}_{13}/\chi^{(3)}_{14}$ and $d = \chi^{(3)}_{12} /\chi^{(3)}_{14}$. An interesting prediction is that when the incoming radiations are linearly polarized, the outgoing $s'_3$ will be zero. That occurs because the last six components of the triple Stokes vector are zero for linearly polarized incoming light (see Eq.~\ref{eq:TripleStokes}), and the first ten elements of the last row in $\mathcal{M}^{(3)}$ are zero for real-valued susceptibilities (see Eq.~\ref{eq:TripleMuellerRatio}). The condition of real-valued susceptibilities is further explored in two more specific cases.
\newline
\newline
\textit{1. Triple Mueller Matrix for Isotropic Media}
\newline
For isotropic material with real susceptibilities, only three susceptibility components are nonzero, of which only one is independent: $\chi^{(3)}_{11}=\chi^{(3)}_{22}=3\chi^{(3)}_{14}$~\cite{Boyd2008_Nonli}. By substituting these symmetry relations into Eq.~\ref{eq:TripleMuellerRatio}, the matrix for isotropic material is obtained as follows:
\onecolumngrid
%%%% Eq: Tripel Stokes Poloarimetry  %%%%%
\begin{equation}\label{eq:MuellerIsotropic}
\arraycolsep=4 pt
M^{(3)}_\text{iso} = \small
{(\chi^{(3)}_{14})}^2 \,
\left(\begin{array}{cccccccccccccccc} 5\, \sqrt{2} & \frac{4\, \sqrt{6}}{3} & \frac{8\, \sqrt{3}}{3} & 0 & 0 & 3 & 0 & 0 & 0 & 3 & 0 & 0 & 0 & 0 & 0 & 0\\ 0 & - \frac{\sqrt{6}}{3} & \frac{\sqrt{3}}{3} & 9 & 0 & -3 & 0 & 0 & 0 & 3 & 0 & 0 & 0 & 0 & 0 & 0\\ 0 & 0 & 0 & 0 & 9 & 0 & 1 & 3 & 3 & 0 & 0 & 0 & 0 & 0 & 0 & 0\\ 0 & 0 & 0 & 0 & 0 & 0 & 0 & 0 & 0 & 0 & 9 & 0 & -1 & -3 & 3 & 0 \end{array}\right)
\end{equation}

\twocolumngrid

As can be seen, and may readily be verified by an experiment, the matrix elements for isotropic material scale only with an effective susceptibility value.
\newline
\newline
\textit{2. Triple Mueller Matrix for Hexagonal Symmetry}

For materials possessing hexagonal and Kleinman symmetry, three elements are independent: $\chi_{11}^{(3)}, \, \chi_{22}^{(3)}$ and $\chi_{14}^{(3)}$. Substituting them in Eq.~\ref{eq:TripleMuellerRatio}, the hexagonal material matrix assumes the form expressed in Eq.~\ref{eq:TripleMuellerHexagonal}

Eq.~\ref{eq:TripleMuellerHexagonal} assumes that the cylindrical axis is oriented along the axis $i=1$. The experimentally-measured Mueller matrix can be used to extract the susceptibility values of the material. For example, the cylindrically symmetric crystalline aggregates of biologically important molecules such as astaxsanthin and $\beta$-carotene have been characterized with THG polarimetric microscopy and $\chi^{(3)}_{11}/\chi^{(3)}_{14}$ and $\chi^{(3)}_{22}/\chi^{(3)}_{14}$ were extracted from the measurements~\cite{Tokarz2014_Molec}.
\onecolumngrid
%%%% Eq: Tripel Stokes Poloarimetry  %%%%%
\begin{equation}\label{eq:TripleMuellerHexagonal}
\begin{split}
& M^{(3)}_\text{hex} = {(\chi^{(3)}_{14})}^2
\footnotesize
\arraycolsep=2 pt
\medmuskip = 0.5mu % default: 4mu plus 2mu minus 4mu
\left(\begin{array}{cccccccccccccccc} 
\frac{\sqrt{2}}{4} \left(a^2 + b^2 + 2\right) & \frac{\sqrt{6}}{12} \left(a^2 + b^2 - 2\right) & \frac{\sqrt{3}}{6} \left(a^2 + b^2 - 2\right) & \frac{1}{2}(a^2 - b^2) & 0 & b & 0 & 0 & 0 & a & 0 & 0 & 0 & 0 & 0 & 0\\ 
\frac{\sqrt{2}}{4} \left(a^2 - b^2\right) & \frac{\sqrt{6}}{12} \left( a^2 - b^2 - 4\right) & \frac{\sqrt{3}}{6} \left(a^2 - b^2 + 2\right) & \frac{1}{2}(a^2 + b^2) & 0 & -b & 0 & 0 & 0 &  a & 0 & 0 & 0 & 0 & 0 & 0\\ 
0 & 0 & 0 & 0 & a\, b & 0 & 1 & b & a & 0 & 0 & 0 & 0 & 0 & 0 & 0\\ 
0 & 0 & 0 & 0 & 0 & 0 & 0 & 0 & 0 & 0 & a\, b & 0 & -1 & -b & a & 0 \end{array}\right)
\end{split}
\end{equation}
\twocolumngrid
\section{Extraction of Third-order Susceptibilities and Phases from the Mueller Matrix}\label{sec:Chis}
Following the derivations analogous to the two-photon and general Stokes Mueller polarimetry~\cite{Azzam1977_Ellip,Anderson1994_Neces,Samim2015_Doubl,Samim2015_NLSM}, the expression for third-order susceptibility tensor elements can be obtained from the triple Mueller matrix elements. In deriving the expression for products of a pair of the susceptibility tensor elements, first matrices $\mathcal{T}$ and $\Gamma$ are derived by vectorizing the Pauli $\tau$ and $\gamma$ matrices, respectively (see Eqs.~\ref{eq:T} and~\ref{eq:Gamma} in Appendix~\ref{sec:gamma}). Both $\mathcal{T}$ and $\Gamma$ matrices are invertible and obey $\mathcal{T}^{-1}=\textstyle{1 \over 2} \mathcal{T}^\dag$, $\Gamma^{-1}=\textstyle{1 \over 2} \Gamma^\dag$, respectively.

Next the pair-wise products of the susceptibilities can be obtained as:
%%%% Eq: Tripel Stokes Poloarimetry  %%%%%
\begin{equation}\label{eq:X}
\rm{X}^{(3)}= \mathcal{T}^{-1}\mathcal{M}^{(3)}\Gamma
\end{equation}
where the matrix $\rm{X}^{(3)}=< \chi^{(3)} \otimes \chi^{(3)*}>_e $ is for the ensemble average, as was shown in Eq.~\ref{eq:MuellerNLEnsemble2}.

In the elemental form, the susceptibility products can be found using $\text{X}_{ij}=\frac{1}{2}\mathcal{T}^\dag _{it}M_{t N}\Gamma_{Nj}$, where $i = (a-1)2+b$ and $j =(A-1)4+B$. Since, $\chi_{aA}^{}\chi_{bB}^*=|\chi_{aA}||\chi_{bB}|e^{i(\delta_{aA}-\delta_{bB})}$, then the relative phase between any two susceptibility elements ($\delta_{aA}^{}-\delta_{bB}^{}$) in the case of a non-depolarizing sample can be found according to:
\onecolumngrid
\begin{equation}\label{eq:RelPhase3}
\begin{split}
&\delta_{aA}-\delta_{aA} 
=\Delta_{aA,bB} = \tan^{-1}\left(-\mathrm{i}\frac{\chi_{aA}^{}\chi_{bB}^*-\chi_{bB}^{}\chi_{aA}^*}{\chi_{aA}^{}\chi_{bB}^*+\chi_{bB}^{}\chi_{aA}^*}\right)
=\tan^{-1}\left(\mathrm{i}\frac{\text{X}_{kl}-\text{X}_{ij}}{\text{X}_{kl}+\text{X}_{ij}}\right) = \tan^{-1}\left(\mathrm{i}\frac{\mathcal{T}^\dag _{kt}M_{t N}\Gamma_{Nl}-\mathcal{T}^\dag _{it}M_{t N}\Gamma_{Nj}}{\mathcal{T}^\dag _{kt}M_{t N}\Gamma_{Nl}+\mathcal{T}^\dag _{it}M_{tN}\Gamma_{Nj}}\right)  
\end{split}
\end{equation}
\twocolumngrid
where $k = (b-1)2+a$ and $l =(B-1)4+A$, and summations is performed over repeated indices. Equations~\ref{eq:X} and~\ref{eq:RelPhase3} show that the products of the susceptibility tensor components and the relative phases between the components can be obtained from the Mueller matrix. In the next section, the description of polarimetric measurements will be presented to obtain the Mueller matrix and $\chi^{(3)}$ tensor values.
\section{Measurement of Triple Mueller Matrix of the Medium by THG Polarimetry}\label{sec:THGMeasurementPol}
\begin{figure}[htbs]
	\includegraphics[width=0.48\textwidth]{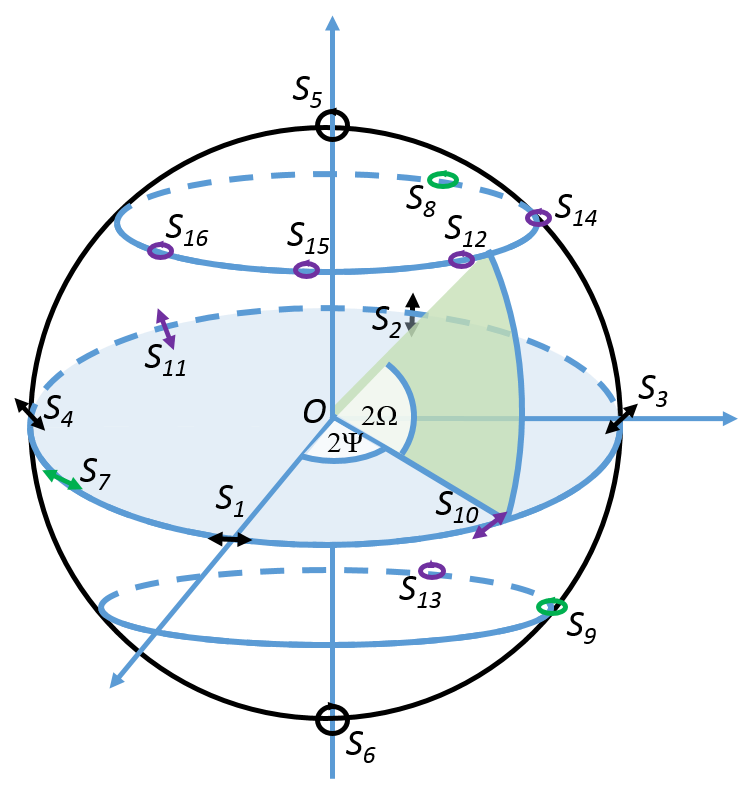}
	\caption[Poincar\'{e} sphere for triple Stokes vector.]{Poincar\'e sphere for tiple Stokes vector. States $S_1$...$S_{16}$ have the respective Poincar\'e sphere coordinate values: $(\Psi,\Omega) = (
		\left[0,0\right], \left[\frac{\pi}{2},0\right],\left[\frac{\pi}{4},0\right],\left[- \frac{\pi}{4},0\right],\left[0,\frac{\pi}{4}\right],
		\left[0,- \frac{\pi}{4}\right]$, $\left[- \frac{\pi}{8},0\right],
		\left[\frac{\pi}{2}\frac{\pi}{8}\right], \left[\frac{\pi}{4},- \frac{\pi}{8}\right],
		\left[\frac{\pi}{8},0\right], \left[\frac{3\, \pi}{8},0\right],\left[\frac{\pi}{8},
		\frac{\pi}{8}\right],\left[\frac{\pi}{2},- \frac{\pi}{8}\right]$, $
		\left[\frac{\pi}{4},\frac{\pi}{8}\right], \left[0,\frac{\pi}{8}\right]$, and $\left[- \frac{\pi}{8},\frac{\pi}{8}\right]$). Note $S_1$ to $S_6$ states are used for linear polarimetry, and $S_1$ to $S_9$ states are also used for double Stokes vector~\cite{Samim2015_Doubl}.}\label{fig:Poincare}
\end{figure}

In order to find each element of $\mathcal{M}^{(3)}$ matrix, the outgoing Stokes vector is measured for each of the sixteen unique incoming polarization states. For each measurement $Q$, all four components of the outgoing Stokes vector $s'$ have to be recorded. The solution to the third-order Mueller matrix from the polarimetry data is:
\begin{equation}\label{eq:MuellerInverseS}
\mathcal{M}^{(3)} = s'S^{-1}
\end{equation}
where $s'$ is a measured $4\times16$ matrix containing outgoing Stokes states with components $s'_0$, $s'_1$, $s'_2$, $s'_3$ in columns for the sixteen selected incoming polarization states. $S^{-1}$ is a $16\times 16$ matrix obtained by inverting the matrix for sixteen different incoming polarization states. It is necessary to choose unique incoming polarization states that produce an invertible matrix. A set of prepared polarization states composed of sixteen different orientations for an incoming radiation that generates an invertible matrix $S$ is shown in Fig.~\ref{fig:Poincare}. In terms of Poincar\'e coordinates, the states are as follows: $(\Psi,\Omega) =$ ($
\left[0,0\right],\left[\frac{\pi}{2},0\right],
\left[\frac{\pi}{4},0\right],\left[- \frac{\pi}{4},0\right],\left[0,\frac{\pi}{4}\right],
\left[0,- \frac{\pi}{4}\right],\left[- \frac{\pi}{8},0\right]$, $
\left[\frac{\pi}{2},\frac{\pi}{8}\right],\left[\frac{\pi}{4},- \frac{\pi}{8}\right],
\left[\frac{\pi}{8},0\right],\left[\frac{3\, \pi}{8},0\right],\left[\frac{\pi}{8},
\frac{\pi}{8}\right],\left[\frac{\pi}{2},- \frac{\pi}{8}\right],
\left[\frac{\pi}{4},\frac{\pi}{8}\right]$, $\left[0,\frac{\pi}{8}\right]$, and $\left[- \frac{\pi}{8},\frac{\pi}{8}\right]$). Additionally the $S$ matrix expression is given in Eq.~\ref{eq:S3Matrix} of Appendix~\ref{sec:Poincare}. 

The Mueller matrix obtained from the measurements can be used to extract the $\chi^{(3)}$ values and retardance phases using Eq.~\ref{eq:X} and~\ref{eq:RelPhase3}, respectively. The extracted $\chi^{(3)}$ values are in the laboratory coordinate frame, and they can be used further to extract the molecular susceptibilities of the structure.
%%%%% Fig: Poincare Sphere %%%%%
\subsection*{Reduced THG polarimetry with Linearly Polarized States}\label{sec:THGPIPO}
For structures with real susceptibilities, the susceptibility component values can be obtained by using only the linear incoming and outgoing polarization states. If linear incoming polarization state is used, i.e. $s_3=0$, then the last six components of the $S^{(3)}$ vector are zero (Eq.~\ref{eq:TripleStokes}). Therefore, the outgoing $s'_3$ will also be zero for the structure with real-valued susceptibilities (see Eqs.~\ref{eq:NLStokeMueller},~\ref{eq:TripleStokes} and~\ref{eq:TripleMuellerRatio}). Thus, the reduced Stokes Mueller polarimetry can be employed for measuring real-valued susceptibility components by conducting the so-called polarization-in polarization-out (PIPO) measurements, where linear polarizations of incoming and outgoing radiation are used. PIPO has been shown to provide a robust determination of the laboratory coordinate susceptibility ratios, the molecular susceptibility ratios, and the orientation angle for cylindrically symmetric materials~\cite{Samim2014_Secon,Tuer2012_Hiera,Tokarz2014_Molec}. The PIPO setup uses a linear polarizer (at angle $\theta$) for the fundamental radiation, a nonlinear interaction medium, and a linear polarizer (analyzer at an angle $\varphi$) for the THG. The THG intensity is measured at different polarizer and analyzer orientations, and the surface plot of intensities is constructed as a function of polarizer and analyzer angles.

The dependence of the outgoing THG Stokes vector on the incoming linear polarization orientation can be derived by multiplying the Mueller matrix from Eq.~\ref{eq:TripleMuellerRatio} with the triple Stokes vector of linear polarization (see Appendix~\ref{sec:Poincare} Eq.~\ref{eq:LinTreiplStokes}):
%%%%% Eq: Stokes Poloarimetry  %%%%%
\begin{equation}\label{eq:THGStokes}
\left( {\begin{array}{*{20}{c}}
	{{{s'}_0}\left( {3\omega } \right)}\\
	{{{s'}_1}\left( {3\omega } \right)}\\
	{{{s'}_2}\left( {3\omega } \right)}\\
	{{{s'}_3}\left( {3\omega } \right)}
	\end{array}} \right) \propto \left( {\begin{array}{*{20}{c}}
	{\sigma _1^2 + \sigma _2^2}\\
	{\sigma _1^2 - \sigma _2^2}\\
	{2{\sigma _1}{\sigma _2}}\\
	0
	\end{array}} \right)
\end{equation}
where
%%%%% Eq: Stokes Poloarimetry  %%%%%
\begin{equation}\label{eq:StokeMueller}
\footnotesize
\begin{split}
{\sigma _1}
&= \chi^{(3)}_{12}\, {\cos\!\left(\theta{}\right)}^3 + 3\, \chi^{(3)}_{14}\, {\cos\!\left(\theta{}\right)}^2\, \sin\!\left(\theta{}\right) \\
&+ 3\, \chi^{(3)}_{13}\, \cos\!\left(\theta{}\right)\, {\sin\!\left(\theta{}\right)}^2 + \chi^{(3)}_{11}\, {\sin\!\left(\theta{}\right)}^3 , \\
{\sigma _2} 
&= \chi^{(3)}_{22}\, {\cos\!\left(\theta{}\right)}^3 + 3\, \chi^{(3)}_{24}\, {\cos\!\left(\theta{}\right)}^2\, \sin\!\left(\theta{}\right) \\
&+ 3\, \chi^{(3)}_{23}\, \cos\!\left(\theta{}\right)\, {\sin\!\left(\theta{}\right)}^2 + \chi^{(3)}_{21}\, {\sin\!\left(\theta{}\right)}^3 
\end{split}
\end{equation}
The resulting THG from the nonlinear medium passing through a linear analyzer is:
%%%%% Eq: Stokes Poloarimetry  %%%%%
\begin{equation}\label{eq:TipleStokes}
\begin{split}
s'\left( {3\omega } \right) 
&= {M_{analyzer}}\mathcal{M}^{(3)}S^{(3)}(\theta)\\
&= \mathcal{L} \left( {\begin{array}{*{20}{c}}
	{{{\left( {{\sigma _1}\sin \varphi  + {\sigma _2}\cos \varphi } \right)}^2}}\\
	{\cos (2\varphi ){{\left( {{\sigma _1}\sin \varphi  + {\sigma _2}\cos \varphi } \right)}^2}}\\
	{\sin(2\varphi ){{\left( {{\sigma _1}\sin \varphi  + {\sigma _2}\cos \varphi } \right)}^2}}\\
	0
	\end{array}} \right)
\end{split}
\end{equation}
where $\mathcal{L}$ is a scaling constant accounting for the experimental conditions and is proportional to the intensity of fundamental radiation. From Eq.~\ref{eq:TipleStokes} it follows that for real susceptibilities and linear incoming polarization $s_3'=0$. Therefore, it is very informative to measure the $s_3'$ component, and if the measured value is negligible, real susceptibilities may be assumed for the material. Additionally, $s'_3=0$ shows that the fundamental and THG radiations do not experience birefringence. Such assumption applies often when measuring thin samples at the wavelength away from the fundamental and THG absorption bands. The $s_0'$ component, expressed in Eq.~\ref{eq:TipleStokes}, is similar to the PIPO equation for THG, which has been used previously to investigate crystalline structures~\cite{Tokarz2014_Molec,Tokarz2014_Organ}. It can be used in nonlinear microscopy to fit the PIPO surface plot of THG imaging data, and is more explicitly stated as follows:
%%%%% Eq: Stokes Poloarimetry  %%%%%
\begin{equation}\label{eq:StokeMuellerAnalyzer}
{s'_0}\left( {3\omega } \right) = {I_{3\omega }}(\theta ,\varphi ) = \mathcal{L} \left| {{\sigma _1}\sin \varphi } \right. + {\sigma _2}{\left. {\cos \varphi } \right|^2}
\end{equation}
For a sample with isotropic symmetry, there is only one independent susceptibility ($\chi^{(3)}_{11}=\chi^{(3)}_{22}=3\chi^{(3)}_{14}$):
\begin{equation}\label{eq:THGIsotropic}
\begin{split}
{s'_0({3\omega})}_{iso} 
&\propto  {\left| {\chi^{(3)}_{11}} \right|^2}{\left| {\cos \left( \varphi  \right){\mkern 1mu} \cos \left( \theta  \right) + \sin \left( \varphi  \right){\mkern 1mu} \sin \left( \theta  \right)} \right|^2} \\
&\propto {\left| \cos \left( \varphi -\theta \right)\right|^2}
\end{split}
\end{equation}
Thus, for isotropically symmetric material the outgoing intensity directly depends only on the direction of incoming radiation and scales according to an effective susceptibility. This predication can be tested in an experiment. The isotropic distribution of retinal molecules in the fruit fly eye and the astaxanthin molecules in red aplanospores of \textit{Haematococcus pluvialis} has been observed with THG polarimetric microscopy~\cite{Tokarz2014_Molec}.

For hexagonally symmetric samples, where Kleinman symmetry is also valid, there are three independent nonzero susceptibilities ($\chi_{11}^{(3)}, \chi_{22}^{(3)}$ and $\chi_{14}^{(3)}$):
\begin{equation}
\begin{split}
&{s'_0({3\omega })}_{hex}\propto\\
&\small  
\begin{split} 
&|{\cos \left( \varphi  \right){\mkern 1mu} \left( {{\chi^{(3)}_{22}} \cos {{\left( \theta  \right)}^3} + 3{\chi^{(3)}_{14}} \cos \left( \theta  \right){\mkern 1mu} \sin {{\left( \theta  \right)}^2}} \right) +} \\
& {\sin \left( \varphi  \right){\mkern 1mu} \left( {3{\chi^{(3)}_{14}} \cos {{\left( \theta  \right)}^2}{\mkern 1mu} \sin \left( \theta  \right) + {\chi^{(3)}_{11}} \sin {{\left( \theta  \right)}^3}} \right)}|^2
\end{split}
\end{split}
\end{equation}
This relation can be used to fit the two-dimensional $(\theta,\varphi)$ intensity (PIPO) surface plot when performing polarimetry for a sample possessing hexagonal symmetry such as H-aggregates of astaxanthin and $\beta$-carotene aggregates in orange carrot root cells~\cite{Tokarz2014_Molec,Tokarz2014_Organ}.

\section{Discussion and Conclusion}
The three-photon polarimetry equations, just like in two-photon polarimetry, follows from the general Stokes Mueller polarimetry formalism~\cite{Samim2015_NLSM}. The dimensions of coherency matrix, the polarization state vector and the material matrix are larger for three-photon compared to the two-photon polarimetry. The $4\times4$ $\gamma$ matrices expand the coherency matrix as well as the corresponding Mueller matrix. The Mueller matrix $\mathcal{M}^{(3)}$ and polarization state vector $S^{(3)}$ have $4\times16$ and $16\times1$ dimensions, respectively. Similar to two-photon polarimetry, the $\mathcal{M}^{(3)}$- and $\rm X^{(3)}$-matrix components can be sorted into four groups ($NP$, $I$, $O$ and $OI$) with distinct phase relations between the incoming and outgoing retardance effects.

A complete three-photon polarimetry experiment utilizes a set of sixteen unique polarization states of incoming radiation each composed of sixteen triple Stokes vector elements, which forms an invertible matrix, with numerical values given in Eq.~\ref{eq:S3Matrix} in Appendix~\ref{sec:Poincare}. This $16\times 16$ matrix is used together with the outgoing polarization states $s'$ to determine uniquely all elements of the material matrix $\mathcal{M}^{(3)}$. For the material with real susceptibilities, and when Kleinman symmetry is valid, the Mueller matrix is composed of five independent susceptibility tensor elements, and it reduces to four ratios. Therefore, all 64 elements of matrix $\mathcal{M}^{(3)}$ are not independent. In addition, by a reduced polarimetry a subset of $\mathcal{M}^{(3)}$ components may be exploited to determine the susceptibility ratios for certain material symmetries. For instance, PIPO measurements, with linearly polarized states of incoming and outgoing radiations, can be used to deduce the real-valued susceptibilities~\cite{Tokarz2014_Molec,Tokarz2014_Organ}. 

For the isotropically symmetric materials, the Mueller matrix is composed of constants, independent of susceptibilities, and the THG signal scales with an effective susceptibility (Eq.~\ref{eq:MuellerIsotropic}). Thus, the outgoing radiation has a simple dependence on the $\mathcal{M}^{(3)}$ matrix, and the matrix component values can be easily verified experimentally with the polarimetric measurement in isotropic media. The triple Mueller matrix of hexagonally symmetric material depends on two ratios, and some elements depend only on a single ratio. Therefore, in designing experiments for a material possessing hexagonal symmetry, a reduced polarimetry may be conducted to obtain the susceptibility component ratios, using only a few polarization states of incoming and outgoing radiation.

In summary, we presented a framework for the triple Stokes Mueller polarimetry of three-photon processes. The theory is provided in the context and in analogy with previous works on Stokes Mueller formalism as well as conventional nonlinear optics notations~\cite{Samim2015_Doubl,Samim2015_NLSM}. The derived equations relate the outgoing Stokes vector for the THG signal to the polarization state of the incoming fundamental radiation beam and to the third-order susceptibility tensor values of the intervening medium. We have additionally described the method for performing a complete three-photon Stokes Mueller polarimetry, which requires sixteen independent polarization states for the incoming radiation. Various symmetries of a material can be explored with expressions of triple Mueller matrix, and consequently, a reduced polarimetry may be performed for known symmetries to extract the corresponding susceptibility ratios and orientations of the principle axis of the material.

The triple Stokes Mueller polarimetry can be extended to other three-photon processes as well, including CARS, by redefining the polarization state vector for incoming radiation. However, the validity of the intensity independent susceptibilities has to be tested by performing experiments with different incoming radiation intensities, especially when the incoming radiation wavelength approaches resonant transitions of the molecules in the material. The extension of three-photon polarimetry to other nonlinear processes provide a great opportunity for future research into triple Stokes Mueller polarimetry.
\section*{Appendix}
\appendix
\twocolumngrid
\section{Pauli and Gamma Matrices}\label{sec:gamma}
Three-photon polarimetry requires the expansion of the coherency matrix for the incoming and outgoing radiation to obtain the corresponding real-valued Stokes vector components, as well as the triple Mueller matrix elements. Due to a size difference between the dimension of the matrices for the incoming and outgoing radiations, the corresponding expansions are also different. The outgoing radiation is expanded by the conventional $2\times 2$ matrices, also known as Pauli matrices, while the incoming radiation and $\mathcal{M}^{(3)}$ require $4\times4$ matrices.

The Pauli $\tau$ matrices and $2\times 2$ identity matrix are special case of $\eta$ matrices with dimension 2~\cite{Samim2015_NLSM}. They have the orthogonal property $\text{Tr}(\tau_\mu\tau_\nu)=2\delta_{\mu\nu}$, where $\delta_{\mu\nu}$ is the Kronecker delta. 
%%%%% Eq: Pauli Matrices  %%%%%
\begin{equation}\label{eq:PauliXZ}
\begin{array}{*{20}{c}}
{{\tau}_0 = \left( {\begin{array}{*{20}{c}}
		1&0\\
		0&1
		\end{array}} \right)} & {{\tau}_1 = \left( {\begin{array}{*{20}{c}}
		1&0\\
		0&-1
		\end{array}} \right)}\\
{{\tau}_2 = \left( {\begin{array}{*{20}{c}}
		0&1\\
		1&0
		\end{array}} \right)}&{{\tau}_3 = \left( {\begin{array}{*{20}{c}}
		0&-i\\
		i&0
		\end{array}} \right)}
\end{array}
\end{equation}
The matrix $\mathcal{T}$, used in Section~\ref{sec:Chis} to derive the susceptibility tensor elements in terms of the Mueller matrix, is obtained by the vectorization operation, where each row of the matrix $\mathcal{T}$ comes from the Pauli matrices. $\mathcal{T}$ is invertible and obeys $\mathcal{T}^{-1}=\textstyle{1 \over 2} \mathcal{T}^\dag$:
\begin{equation}\label{eq:T}
\mathcal{T}
\equiv
\left(
\begin{matrix}
\text{vec}(\tau_0),
\cdots,
\text{vec}(\tau_3)
\end{matrix}
\right)^T 
=
\left(
\begin{matrix}
1&0&0&1\\
1&0&0&-1\\
0&1&1&0\\
0&\mathrm{i}&-\mathrm{i}&0\\
\end{matrix}
\right)
\end{equation}
Following the recipe as described in ref.~\cite{Samim2015_NLSM}, the $\gamma$ matrices for three-photon polarimetry are developed in two steps: First, the matrix $\gamma''_{jk}$ is defined such that only the value of element $jk$ of the matrix $\gamma''_{jk}$ is $1$, and $0$ for all other elements (both $j$ and $k$ run from 1 to $4$). This creates a two dimensional set of matrices, where each element of the set is a $4\times 4$ matrix.
\newline
\newline
\onecolumngrid
\begin{equation}
\arraycolsep=4pt\def\arraystretch{0.8}
\begin{matrix}
\gamma''_{1,1} =
\begin{pmatrix}
1&0&0&0\\
0&0&0&0\\
0&0&0&0\\
0&0&0&0\\
\end{pmatrix}
&\gamma''_{1,2} =
\begin{pmatrix}
0&1&0&0\\
0&0&0&0\\
0&0&0&0\\
0&0&0&0\\
\end{pmatrix}
&\gamma''_{1,3} =
\begin{pmatrix}
0&0&1&0\\
0&0&0&0\\
0&0&0&0\\
0&0&0&0\\
\end{pmatrix}
&\gamma''_{1,4} =
\begin{pmatrix}
0&0&0&1\\
0&0&0&0\\
0&0&0&0\\
0&0&0&0\\
\end{pmatrix}\\
\gamma''_{2,1} =
\begin{pmatrix}
0&0&0&0\\
1&0&0&0\\
0&0&0&0\\
0&0&0&0\\
\end{pmatrix}
&\gamma''_{2,2} =
\begin{pmatrix}
0&0&0&0\\
0&1&0&0\\
0&0&0&0\\
0&0&0&0\\
\end{pmatrix}
&\gamma''_{2,3} =
\begin{pmatrix}
0&0&0&0\\
0&0&1&0\\
0&0&0&0\\
0&0&0&0\\
\end{pmatrix}
&\gamma''_{2,4} =
\begin{pmatrix}
0&0&0&0\\
0&0&0&1\\
0&0&0&0\\
0&0&0&0\\
\end{pmatrix}\\
\gamma''_{3,1} =
\begin{pmatrix}
0&0&0&0\\
0&0&0&0\\
1&0&0&0\\
0&0&0&0\\
\end{pmatrix}
&\gamma''_{3,2} =
\begin{pmatrix}
0&0&0&0\\
0&0&0&0\\
0&1&0&0\\
0&0&0&0\\
\end{pmatrix}
&\gamma''_{3,3} =
\begin{pmatrix}
0&0&0&0\\
0&0&0&0\\
0&0&1&0\\
0&0&0&0\\
\end{pmatrix}
&\gamma''_{3,4} =
\begin{pmatrix}
0&0&0&0\\
0&0&0&0\\
0&0&0&1\\
0&0&0&0\\
\end{pmatrix}\\
\gamma''_{4,1} =
\begin{pmatrix}
0&0&0&0\\
0&0&0&0\\
0&0&0&0\\
1&0&0&0\\
\end{pmatrix}
&\gamma''_{4,2} =
\begin{pmatrix}
0&0&0&0\\
0&0&0&0\\
0&0&0&0\\
0&1&0&0\\
\end{pmatrix}
&\gamma''_{4,3} =
\begin{pmatrix}
0&0&0&0\\
0&0&0&0\\
0&0&0&0\\
0&0&1&0\\
\end{pmatrix}
&\gamma''_{4,4} =
\begin{pmatrix}
0&0&0&0\\
0&0&0&0\\
0&0&0&0\\
0&0&0&1\\
\end{pmatrix}\\
\end{matrix}
\end{equation}
\twocolumngrid
Note that $\gamma''$ matrices can also expand the coherency matrix $\rho^{(3)}$ of Eq.~\ref{eq:TripleCoherency}. However, the resulting vector for $S^{(3)}$ and $\mathcal{M}^{(3)}$ matrix will be complex. In a second step, in order to obtain the desired hermitian matrices for a third-order process, the following relations are used to define the two-dimensional $\gamma'$ matrices:
\begin{equation}\label{eq:GGM}
\gamma'_{jk}=
\begin{dcases}
\gamma''_{jk}+\gamma''_{kj} \; ,   \hskip 70pt \text{if } j<k\\
\mathrm{i}(\gamma''_{jk}-\gamma''_{kj})\; ,   \hskip 60pt \text{if } j>k\\
\sqrt{\frac{2}{j^2+j}} \left[ \left(\sum\limits_{m=1}^{j}\gamma''_{mm}\right) -j\gamma''_{j+1,j+1} \right]\; ,    \\ \hskip 80pt \text{if } 1\leq k=j<4\\
{1\over \sqrt{2}}\mathcal{I}_{4}\; ,    \hskip 65pt \text{if } j=k=4
\end{dcases}
\end{equation}
where $\mathcal{I}_{4}$ is the $4\times 4$ identity matrix. For the first case (when $j<k$),  matrices $\gamma'_{jk}=\gamma''_{jk}+\gamma''_{kj}$ are real-valued; for the second case (when $j>k$), matrices $\gamma'_{jk}=\mathrm{i}(\gamma''_{jk}-\gamma''_{kj})$ are complex-valued and have similar nonzero elements as in their real-value counterparts obtained in the first case. In the third case, (when $1\leq j=k< 4$), the matrices are diagonal and real-valued. In the last case the identity matrix is used.

Finally, the two-dimensional $\gamma'$ set is converted to a one-dimensional set of matrices~\footnote{This is to simplify the indices and to conform to a Stokes Mueller notation of vector = matrix $\times$ vector. The matrices $\gamma'$ can also be used directly for polarimetry, in which case there will be an additional index for the entity representing the incoming radiation as well as for the entity representing the medium.}: $\gamma'_{jk}\rightarrow\gamma^{}_N$, where $N = 1,\cdots,16$.
In Eq.~\ref{eq:gamma}, the two dimensional $4\times 4$ $\gamma'$ set is shown but labeled with the one-dimensional sixteen-element set $\gamma$. The presented order of $\gamma$ matrices in Eq.~\ref{eq:gamma} is chosen to remain consistent with the order of the linear Stokes vector (determined by Pauli matrices in Eq.~\ref{eq:PauliXZ}) as well as the second-order Stokes vector (determined by Gell-Mann matrices in~\cite{Samim2015_Doubl}). These matrices satisfy all the requirements as desired for expanding the coherency matrix for the nonlinear polarimetry. In addition, the new matrices defined in Eq.~\ref{eq:GGM} ensure that $\gamma$ obey: $\text{Tr}(\gamma_M\gamma_N)=2\delta_{MN}$.
\onecolumngrid
\begin{equation}\label{eq:gamma}
\arraycolsep=2.3pt\def\arraystretch{0.8}
\begin{array}{cccc} 
\gamma_{4\;}=\left(\begin{array}{cccc} 
1 & 0 & 0 & 0\\
0 & -1 & 0 & 0\\ 
0 & 0 & 0 & 0\\ 
0 & 0 & 0 & 0 
\end{array}\right) 
&\quad\gamma_{5}= \left(\begin{array}{cccc} 
0 & 1 & 0 & 0\\
1 & 0 & 0 & 0\\
0 & 0 & 0 & 0\\ 
0 & 0 & 0 & 0 
\end{array}\right) 
&\quad \gamma_{9}=\left(\begin{array}{cccc} 
0 & 0 & 1 & 0\\ 
0 & 0 & 0 & 0\\ 
1 & 0 & 0 & 0\\ 
0 & 0 & 0 & 0 
\end{array}\right) 
&\quad \gamma_{10}=\left(\begin{array}{cccc} 
0 & 0 & 0 & 1\\ 
0 & 0 & 0 & 0\\ 
0 & 0 & 0 & 0\\ 
1 & 0 & 0 & 0 
\end{array}\right)\\ 
\gamma_{11}=\left(\begin{array}{cccc} 
0 & -\mathrm{i} & 0 & 0\\ 
\mathrm{i} & 0 & 0 & 0\\ 
0 & 0 & 0 & 0\\ 
0 & 0 & 0 & 0 
\end{array}\right) 
& \gamma_{3}= \frac{\sqrt{3}}{3}\left(\begin{array}{cccc} 
1 & 0 & 0 & 0\\ 
0 & 1 & 0 & 0\\ 
0 & 0 & - 2 & 0\\ 
0 & 0 & 0 & 0 
\end{array}\right) 
&\quad \gamma_{6\;}=\left(\begin{array}{cccc} 
0 & 0 & 0 & 0\\ 
0 & 0 & 1 & 0\\ 
0 & 1 & 0 & 0\\ 
0 & 0 & 0 & 0 
\end{array}\right) 
&\quad \gamma_{8\;}=\left(\begin{array}{cccc} 
0 & 0 & 0 & 0\\ 
0 & 0 & 0 & 1\\ 
0 & 0 & 0 & 0\\ 
0 & 1 & 0 & 0 
\end{array}\right)\\ 
\gamma_{15}=\left(\begin{array}{cccc} 
0 & 0 & -\mathrm{i} & 0\\ 
0 & 0 & 0 & 0\\ 
\mathrm{i} & 0 & 0 & 0\\ 
0 & 0 & 0 & 0 \end{array}\right) 
&\quad \gamma_{12}=\left(\begin{array}{cccc} 
0 & 0 & 0 & 0\\ 
0 & 0 & -\mathrm{i} & 0\\ 
0 & \mathrm{i} & 0 & 0\\ 
0 & 0 & 0 & 0 
\end{array}\right) 
& \gamma_{2\;}=\frac{\sqrt{6}}{6}\left(\begin{array}{cccc} 
1 & 0 & 0 & 0\\ 
0 & 1 & 0 & 0\\ 
0 & 0 & 1 & 0\\ 
0 & 0 & 0 & - 3 
\end{array}\right) 
&\quad \gamma_{7\;}=\left(\begin{array}{cccc} 
0 & 0 & 0 & 0\\ 0 & 0 & 0 & 0\\
0 & 0 & 0 & 1\\
0 & 0 & 1 & 0 \end{array}\right)\\ 
\gamma_{16}=\left(\begin{array}{cccc} 
0 & 0 & 0 & -\mathrm{i}\\ 
0 & 0 & 0 & 0\\ 
0 & 0 & 0 & 0\\ 
\mathrm{i} & 0 & 0 & 0 \end{array}\right) 
&\quad \gamma_{14}= \left(\begin{array}{cccc} 
0 & 0 & 0 & 0\\ 
0 & 0 & 0 & -\mathrm{i}\\ 
0 & 0 & 0 & 0\\ 
0 & \mathrm{i} & 0 & 0 \end{array}\right) 
&\quad  \gamma_{13}=\left(\begin{array}{cccc} 
0 & 0 & 0 & 0\\ 
0 & 0 & 0 & 0\\ 
0 & 0 & 0 & -\mathrm{i}\\ 
0 & 0 & \mathrm{i} & 0 \end{array}\right) 
&\gamma_{1} = \frac{\sqrt{2}}{2}\left(\begin{array}{cccc} 
1 & 0 & 0 & 0\\ 
0 & 1 & 0 & 0\\ 
0 & 0 & 1 & 0\\ 
0 & 0 & 0 & 1 
\end{array}\right) \end{array}
\end{equation}
\newline
\twocolumngrid
The matrices $\mathcal{T}$ and $\Gamma$ are used in Section~\ref{sec:Chis} to derive the third-order susceptibilities and their phases in terms of the triple Mueller matrix elements. $\Gamma$ is derived from $\gamma$ matrices in Eq.~\ref{eq:gamma}, where each row of $\Gamma$ is expressed by vectorizing $\gamma$ matrices. $\Gamma$ is invertible and obeys $\Gamma^{-1}=\textstyle{1 \over 2} \Gamma^\dag$.
\begin{equation}\label{eq:Gamma}
\arraycolsep=1.0pt
\begin{split}
\Gamma{}
&\equiv
\left(
\begin{matrix}
\text{vec}(\gamma_1),
\text{vec}(\gamma_2),
\cdots,
\text{vec}(\gamma_{15}),
\text{vec}(\gamma_{16})
\end{matrix}
\right)^T \\
&=
\left(\begin{array}{cccccccccccccccc} \frac{\sqrt{2}}{2} & 0 & 0 & 0 & 0 & \frac{\sqrt{2}}{2} & 0 & 0 & 0 & 0 & \frac{\sqrt{2}}{2} & 0 & 0 & 0 & 0 & \frac{\sqrt{2}}{2}\\ \frac{\sqrt{6}}{6} & 0 & 0 & 0 & 0 & \frac{\sqrt{6}}{6} & 0 & 0 & 0 & 0 & \frac{\sqrt{6}}{6} & 0 & 0 & 0 & 0 &  \frac{-3}{\sqrt{6}}\\ \frac{\sqrt{3}}{3} & 0 & 0 & 0 & 0 & \frac{\sqrt{3}}{3} & 0 & 0 & 0 & 0 &  \frac{-2}{\sqrt{3}} & 0 & 0 & 0 & 0 & 0\\ 1 & 0 & 0 & 0 & 0 & -1 & 0 & 0 & 0 & 0 & 0 & 0 & 0 & 0 & 0 & 0\\ 0 & 1 & 0 & 0 & 1 & 0 & 0 & 0 & 0 & 0 & 0 & 0 & 0 & 0 & 0 & 0\\ 0 & 0 & 0 & 0 & 0 & 0 & 1 & 0 & 0 & 1 & 0 & 0 & 0 & 0 & 0 & 0\\ 0 & 0 & 0 & 0 & 0 & 0 & 0 & 0 & 0 & 0 & 0 & 1 & 0 & 0 & 1 & 0\\ 0 & 0 & 0 & 0 & 0 & 0 & 0 & 1 & 0 & 0 & 0 & 0 & 0 & 1 & 0 & 0\\ 0 & 0 & 1 & 0 & 0 & 0 & 0 & 0 & 1 & 0 & 0 & 0 & 0 & 0 & 0 & 0\\ 0 & 0 & 0 & 1 & 0 & 0 & 0 & 0 & 0 & 0 & 0 & 0 & 1 & 0 & 0 & 0\\ 0 & \mathrm{i} & 0 & 0 & -\mathrm{i} & 0 & 0 & 0 & 0 & 0 & 0 & 0 & 0 & 0 & 0 & 0\\ 0 & 0 & 0 & 0 & 0 & 0 & \mathrm{i} & 0 & 0 & -\mathrm{i} & 0 & 0 & 0 & 0 & 0 & 0\\ 0 & 0 & 0 & 0 & 0 & 0 & 0 & 0 & 0 & 0 & 0 & \mathrm{i} & 0 & 0 & -\mathrm{i} & 0\\ 0 & 0 & 0 & 0 & 0 & 0 & 0 & \mathrm{i} & 0 & 0 & 0 & 0 & 0 & -\mathrm{i} & 0 & 0\\ 0 & 0 & \mathrm{i} & 0 & 0 & 0 & 0 & 0 & -\mathrm{i} & 0 & 0 & 0 & 0 & 0 & 0 & 0\\ 0 & 0 & 0 & \mathrm{i} & 0 & 0 & 0 & 0 & 0 & 0 & 0 & 0 & -\mathrm{i} & 0 & 0 & 0 \end{array}\right)
\end{split}
\end{equation}
\section{$S^{(3)}$ Vector for Various Polarizations}\label{sec:Poincare}
\subsection{Triple Stokes Vector for Linearly Polarized States}
When the incoming electric field is linearly polarized at an angle $\theta$ form the primary axis: $E(\omega) = [E_1(\omega),E_2(\omega)]^{T} = E_0 [\sin{\theta},\cos{\theta}]^{T}$. Substituting this in Eq.~\ref{eq:TripleStokes}, $S^{(3)}(\theta)$, representing the linearly polarized incoming radiations is obtained (see Eq.~\ref{eq:LinTreiplStokes}). Multiplying Eq.~\ref{eq:LinTreiplStokes} and Eq.~\ref{eq:TripleMuellerRatio}, the outgoing THG Stokes vector is obtained as shown in Eq.~\ref{eq:THGStokes}. The outgoing Stokes vector shows that by using incoming linearly-polarized states the outgoing polarization will also be linear (i.e. $s'_3=0$ in Eq.~\ref{eq:THGStokes}). This feature is used in linear PIPO measurements as described in Section~\ref{sec:THGPIPO}.
\begin{equation}\label{eq:LinTreiplStokes}
\begin{split}
& S^{(3)}(\theta) = \left<E_0^6\right> \\ 
&\footnotesize{
	\left(\begin{array}{c} \frac{\sqrt{2}}{2} [{\cos\!\left(\theta{}\right)}^6 + 9\, {\cos\!\left(\theta{}\right)}^4\, {\sin\!\left(\theta{}\right)}^2 + 9\,  {\cos\!\left(\theta{}\right)}^2\, {\sin\!\left(\theta{}\right)}^4 + { {\sin\!\left(\theta{}\right)}^6}]\\
	\frac{\sqrt{6}}{6} [{\cos\!\left(\theta{}\right)}^6- 27\, {\cos\!\left(\theta{}\right)}^4\, {\sin\!\left(\theta{}\right)}^2 + 9\, {\cos\!\left(\theta{}\right)}^2\, {\sin\!\left(\theta{}\right)}^4 + {\sin\!\left(\theta{}\right)}^6]\\ 
	\frac{\sqrt{3}}{3} [{\cos\!\left(\theta{}\right)}^6 - 18\,  {\cos\!\left(\theta{}\right)}^2\, {\sin\!\left(\theta{}\right)}^4 + {\sin\!\left(\theta{}\right)}^6]\\
	{\sin\!\left(\theta{}\right)}^6 - {\cos\!\left(\theta{}\right)}^6\\ 
	2\, {\cos\!\left(\theta{}\right)}^3\, {\sin\!\left(\theta{}\right)}^3\\ 6\, {\cos\!\left(\theta{}\right)}^4\, {\sin\!\left(\theta{}\right)}^2\\ 18\, {\cos\!\left(\theta{}\right)}^3\, {\sin\!\left(\theta{}\right)}^3\\ 6\, {\cos\!\left(\theta{}\right)}^5\, \sin\!\left(\theta{}\right)\\ 6\, \cos\!\left(\theta{}\right)\, {\sin\!\left(\theta{}\right)}^5\\ 6\, {\cos\!\left(\theta{}\right)}^2\, {\sin\!\left(\theta{}\right)}^4\\ 0\\ 0\\ 0\\ 0\\ 0\\ 0 \end{array}\right)
}
\end{split}
\end{equation}
\subsection{Triple Stokes Vector on Poincar\'e Sphere}
Poincar\'e sphere is a useful geometrical representation by which different polarization states can be visually identified. In addition, in order to determine values of the Mueller matrix $\mathcal{M}^{(3)}$ elements from a polarimetry measurements, numerical values for the incoming polarization states are required (see Eq.~\ref{eq:MuellerInverseS}). The numerical values can be obtained by deriving the triple Stokes vector in terms of Poincar\'e coordinates in general, and subsequently evaluating triple Stokes vector elements values by substituting the corresponding state's coordinates. The $4\times1$ Stokes vector in the Poincar\'e sphere coordinates is:
\begin{equation}\label{eq:LinearStokePoincare}
s = 
\left<E_0^2\right>
\left(
\begin{matrix}
1\\
\cos(2\Psi)\cos(2\Omega)\\
\sin(2\Psi)\cos(2\Omega)\\
\sin(2\Omega)
\end{matrix}
\right)
\end{equation}
where the variables $2\Psi$ and $2\Omega$ are the azimuth and latitude coordinates on the Poincar\'e sphere as shown in Fig.~\ref{fig:Poincare}. The triple Stokes vector in terms of Poincar\'e coordinates is derived by substituting $s$ values from Eq.~\ref{eq:LinearStokePoincare} into Eq.~\ref{eq:TripleStokes} and is shown in Eq.~\ref{eq:TriplePoincare}.

\subsection{The Matrix of the Prepared Triple Stokes Vector States for Complete Mueller Polarimetry Measurement}
The matrix of prepared triple Stokes states for the complete polarimetry measurements has to be invertible. One invertible matrix can be constructed from the sixteen polarization states given by the Poincar\'e coordinates in Eq.~\ref{eq:TripleStates} and shown on Poincar\'e sphere in Fig.~\ref{fig:Poincare}. By substituting these coordinates into Eq.~\ref{eq:LinearStokePoincare}, the $4\times 1$ Stokes vector values are obtained as in Eq.~\ref{eq:LinearSixteenStates}. Furthermore, by substituting the Poincar\'e coordinate values from Eq.~\ref{eq:TripleStates} into Eq.~\ref{eq:TriplePoincare} below, the triple Stokes vector values for the sixteen states are obtained (see Eq.~\ref{eq:S3Matrix}).
\onecolumngrid
\begin{equation}\label{eq:TriplePoincare}
\small
\begin{split}
& S^{(3)}(\Psi,\Omega) =\textstyle{{1\over4}}\left<{E_0}^6\right> =\scriptsize\left(\begin{array}{c} 
\sqrt{2}[5 - 3{\cos\!\left(2\, \Omega{}\right)}^2\, {\cos\!\left(2\, \Psi{}\right)}^2]\\
\sqrt{6}[- 3 {\cos\!\left(2\, \Omega{}\right)}^3\, {\cos\!\left(2\, \Psi{}\right)}^3 + 2{\cos\!\left(2\, \Omega{}\right)}^2\, {\cos\!\left(2\, \Psi{}\right)}^2 + 3 \cos\!\left(2\, \Omega{}\right)\, \cos\!\left(2\, \Psi{}\right) - \frac{4}{3}]\\ 
\sqrt{3}[3 {\cos\!\left(2\, \Omega{}\right)}^3\, {\cos\!\left(2\, \Psi{}\right)}^3 + 4{\cos\!\left(2\, \Omega{}\right)}^2\, {\cos\!\left(2\, \Psi{}\right)}^2 - 3 \cos\!\left(2\, \Omega{}\right)\, \cos\!\left(2\, \Psi{}\right) - \frac{8}{3}]\\
{\cos\!\left(2\, \Omega{}\right)}^3\, {\cos\!\left(2\, \Psi{}\right)}^3 + 3\, \cos\!\left(2\, \Omega{}\right)\, \cos\!\left(2\, \Psi{}\right)\\ 
{\cos\!\left(2\, \Omega{}\right)}^3\, {\sin\!\left(2\, \Psi{}\right)}^3 - 3\, \cos\!\left(2\, \Omega{}\right)\, {\sin\!\left(2\, \Omega{}\right)}^2\, \sin\!\left(2\, \Psi{}\right)\\ 
3\, \left(\cos\!\left(2\, \Omega{}\right)\, \cos\!\left(2\, \Psi{}\right) - 1\right)\, \left({\cos\!\left(2\, \Omega{}\right)}^2\, \left({\cos\!\left(2\, \Psi{}\right)}^2 - 1\right) - {\cos\!\left(2\, \Omega{}\right)}^2 + 1\right)\\
- 9\, \cos\!\left(2\, \Omega{}\right)\, \sin\!\left(2\, \Psi{}\right)\, \left({\cos\!\left(2\, \Omega{}\right)}^2\, {\cos\!\left(2\, \Psi{}\right)}^2 - 1\right)\\ 
3\, \cos\!\left(2\, \Omega{}\right)\, \sin\!\left(2\, \Psi{}\right)\, {\left(\cos\!\left(2\, \Omega{}\right)\, \cos\!\left(2\, \Psi{}\right) - 1\right)}^2\\
3\, \cos\!\left(2\, \Omega{}\right)\, \sin\!\left(2\, \Psi{}\right)\, {\left(\cos\!\left(2\, \Omega{}\right)\, \cos\!\left(2\, \Psi{}\right) + 1\right)}^2\\ 
- 3\, \left(\cos\!\left(2\, \Omega{}\right)\, \cos\!\left(2\, \Psi{}\right) + 1\right)\, \left({\cos\!\left(2\, \Omega{}\right)}^2\, \left({\cos\!\left(2\, \Psi{}\right)}^2 - 1\right) - {\cos\!\left(2\, \Omega{}\right)}^2 + 1\right)\\ 
-{\sin\!\left(2\, \Omega{}\right)}^3 + 3\, \sin\!\left(2\, \Omega{}\right)\, {\sin\!\left(2\, \Psi{}\right)}^2\, \left({\sin\!\left(2\, \Omega{}\right)}^2 - 1\right)\\ 
6\, \cos\!\left(2\, \Omega{}\right)\, \sin\!\left(2\, \Omega{}\right)\, \sin\!\left(2\, \Psi{}\right)\, \left(\cos\!\left(2\, \Omega{}\right)\, \cos\!\left(2\, \Psi{}\right) - 1\right)\\ 
-9\, \sin\!\left(2\, \Omega{}\right)\, \left(\left({\sin\!\left(2\, \Omega{}\right)}^2 - 1\right)\, \left({\sin\!\left(2\, \Psi{}\right)}^2 - 1\right) - 1\right)\\ 
-3\, \sin\!\left(2\, \Omega{}\right)\, {\left(\cos\!\left(2\, \Omega{}\right)\, \cos\!\left(2\, \Psi{}\right) - 1\right)}^2\\ 
3\, \sin\!\left(2\, \Omega{}\right)\, {\left(\cos\!\left(2\, \Omega{}\right)\, \cos\!\left(2\, \Psi{}\right) + 1\right)}^2\\ 
6\, \cos\!\left(2\, \Omega{}\right)\, \sin\!\left(2\, \Omega{}\right)\, \sin\!\left(2\, \Psi{}\right)\, \left(\cos\!\left(2\, \Omega{}\right)\, \cos\!\left(2\, \Psi{}\right) + 1\right) 
\end{array}\right)
\end{split}
\end{equation}
\twocolumngrid
The $16\times16$ matrix $S$ of the triple Stokes states, given in Eq.~\ref{eq:S3Matrix}, is used in polarimetric measurements to recover the triple Mueller matrix values according to Eq.~\ref{eq:MuellerInverseS}. The chosen triple Stokes states include the states that also are used for linear and double Stokes polarimetry~\cite{Samim2015_Doubl}. However, other states can be used as long as the corresponding $16\times 16$ matrix defined by the triple Stokes polarization states chosen for the measurements matrix is invertible for use in Eq.~\ref{eq:MuellerInverseS}.

\onecolumngrid
\clearpage
\begin{sidewaysfigure}
In terms of Poincar\'{e} coordinates the set for the triple Stokes vector states is (also shown in Fig.~\ref{fig:Poincare}):
\begin{equation}\label{eq:TripleStates}
	\left((\Psi,\Omega)_Q\right)  \quad= \,\,\qquad
	\scriptsize\arraycolsep=1pt\def\arraystretch{0.1}
	\left(\left[0,0\right],\left[\frac{\pi}{2},0\right],\left[\frac{\pi}{4},0\right],\left[- \frac{\pi}{4},0\right],\left[0,\frac{\pi}{4}\right],\left[0,- \frac{\pi}{4}\right],\left[- \frac{\pi}{8},0\right],\left[\frac{\pi}{2},\frac{\pi}{8}\right],\left[\frac{\pi}{4},- \frac{\pi}{8}\right],\left[\frac{\pi}{8},0\right],\left[\frac{3\, \pi}{8},0\right],\left[\frac{\pi}{8},\frac{\pi}{8}\right],\left[\frac{\pi}{2},- \frac{\pi}{8}\right],\left[\frac{\pi}{4},\frac{\pi}{8}\right],\left[0,\frac{\pi}{8}\right],\left[- \frac{\pi}{8},\frac{\pi}{8}\right]\right)\quad
\end{equation}

\begin{equation}\label{eq:LinearSixteenStates}
	\begin{array}{l}
	\quad \left({\bf{s}}^{}_{t,Q}\right) \quad = \quad\quad \,
	\arraycolsep=8.7pt
	\left( {\begin{array}{*{20}{c}}
		{\bf{s}}_{t,1}^{}&{\bf{s}}_{t,2}^{}&{\bf{s}}_{t,3}^{}&{\bf{s}}_{t,4}^{}&{\bf{s}}_{t,5}^{}&{\bf{s}}_{t,6}^{}&{\bf{s}}_{t,7}^{}&{\bf{s}}_{t,8}^{}&{\bf{s}}_{t,9}^{}&{\bf{s}}_{t,10}^{}&{\bf{s}}_{t,11}^{}&{\bf{s}}_{t,12}^{}&{\bf{s}}_{t,13}^{}&{\bf{s}}_{t,14}^{}&{\bf{s}}_{t,15}^{}&{\bf{s}}_{t,16}
		\end{array}} \right)\\
	\left( {\begin{array}{*{20}{c}}
		{{s_{0,Q}}}\\
		{{s_{1,Q}}}\\
		{{s_{2,Q}}}\\
		{{s_{3,Q}}}\\
		\end{array}} \right) \propto \left<{E_0^2}\right> 
		\arraycolsep=9.7pt\def\arraystretch{0.7}
	\left(\begin{array}{cccccccccccccccc} 
	1 & 1 & 1 & 1 & 1 & 1 & 1 & 1 & 1 & 1 & 1 & 1 & 1 & 1 & 1 & 1\\ 1 & -1 & 0 & 0 & 0 & 0 & \frac{\sqrt{2}}{2} & - \frac{\sqrt{2}}{2} & 0 & \frac{\sqrt{2}}{2} & - \frac{\sqrt{2}}{2} & \frac{1}{2} & - \frac{\sqrt{2}}{2} & 0 & \frac{\sqrt{2}}{2} & \frac{1}{2}\\ 0 & 0 & 1 & -1 & 0 & 0 & - \frac{\sqrt{2}}{2} & 0 & \frac{\sqrt{2}}{2} & \frac{\sqrt{2}}{2} & \frac{\sqrt{2}}{2} & \frac{1}{2} & 0 & \frac{\sqrt{2}}{2} & 0 & - \frac{1}{2}\\ 0 & 0 & 0 & 0 & 1 & -1 & 0 & \frac{\sqrt{2}}{2} & - \frac{\sqrt{2}}{2} & 0 & 0 & \frac{\sqrt{2}}{2} & - \frac{\sqrt{2}}{2} & \frac{\sqrt{2}}{2} & \frac{\sqrt{2}}{2} & \frac{\sqrt{2}}{2} 
	\end{array}\right)
	\end{array}
\end{equation}
\begin{equation}\label{eq:S3Matrix}
	\arraycolsep=1.4pt\def\arraystretch{1.9}
	\begin{array}{l}
	%\quad{{\bf{S}}_{N,Q}}\qquad = \quad \left( {\begin{array}{*{20}{c}}
	%	{\bf{S}}_{N,1}^{}&{\bf{S}}_{N,2}^{}&{\bf{S}}_{N,3}^{}&{\bf{S}}_{N,4}^{}&{\bf{S}}_{N,5}^{}&{\bf{S}}_{N,6}^{}&{\bf{S}}_{N,7}^{}&{\bf{S}}_{N,8}^{}&{\bf{S}}_{N,9}^{}&{\bf{S}}_{N,10}^{}&{\bf{S}}_{N,11}^{}&{\bf{S}}_{N,12}^{}&{\bf{S}}_{N,13}^{}&{\bf{S}}_{N,14}^{}&{\bf{S}}_{N,15}^{}&{\bf{S}}_{N,16}
	%	\end{array}} \right)\\
	\left( {\begin{array}{*{20}{c}}
		{{S_{1,Q}}}\\
		{{S_{2,Q}}}\\
		{{S_{3,Q}}}\\
		{{S_{4,Q}}}\\
		{{S_{5,Q}}}\\
		{{S_{6,Q}}}\\
		{{S_{7,Q}}}\\
		{{S_{8,Q}}}\\
		{{S_{9,Q}}}\\
		{{S_{10,Q}}}\\
		{{S_{11,Q}}}\\
		{{S_{12,Q}}}\\
		{{S_{13,Q}}}\\
		{{S_{14,Q}}}\\
		{{S_{15,Q}}}\\
		{{S_{16,Q}}}\\
		\end{array}} \right) \propto \left<{E_0^6}\right> 
	\left(\tiny\arraycolsep=1.1pt\def\arraystretch{3.4}
	\begin{array}{cccccccccccccccc} 
	\frac{\sqrt{2}}{2} & \frac{\sqrt{2}}{2} & \frac{5\, \sqrt{2}}{4} & \frac{5\, \sqrt{2}}{4} & \frac{5\, \sqrt{2}}{4} & \frac{5\, \sqrt{2}}{4} & \frac{7\, \sqrt{2}}{8} & \frac{7\, \sqrt{2}}{8} & \frac{5\, \sqrt{2}}{4} & \frac{7\, \sqrt{2}}{8} & \frac{7\, \sqrt{2}}{8} & \frac{17\, \sqrt{2}}{16} & \frac{7\, \sqrt{2}}{8} & \frac{5\, \sqrt{2}}{4} & \frac{7\, \sqrt{2}}{8} & \frac{17\, \sqrt{2}}{16}\\ \frac{\sqrt{6}}{6} & \frac{\sqrt{6}}{6} & - \frac{\sqrt{6}}{3} & - \frac{\sqrt{6}}{3} & - \frac{\sqrt{6}}{3} & - \frac{\sqrt{6}}{3} & \frac{\sqrt{6}\, \left(9\, \sqrt{2} - 4\right)}{48} & - \frac{\sqrt{6}\, \left(9\, \sqrt{2} + 4\right)}{48} & - \frac{\sqrt{6}}{3} & \frac{\sqrt{6}\, \left(9\, \sqrt{2} - 4\right)}{48} & - \frac{\sqrt{6}\, \left(9\, \sqrt{2} + 4\right)}{48} & \frac{7\, \sqrt{6}}{96} & - \frac{\sqrt{6}\, \left(9\, \sqrt{2} + 4\right)}{48} & - \frac{\sqrt{6}}{3} & \frac{\sqrt{6}\, \left(9\, \sqrt{2} - 4\right)}{48} & \frac{7\, \sqrt{6}}{96}\\ \frac{\sqrt{3}}{3} & \frac{\sqrt{3}}{3} & - \frac{2\, \sqrt{3}}{3} & - \frac{2\, \sqrt{3}}{3} & - \frac{2\, \sqrt{3}}{3} & - \frac{2\, \sqrt{3}}{3} & - \frac{\sqrt{3}\, \left(9\, \sqrt{2} + 8\right)}{48} & \frac{\sqrt{3}\, \left(9\, \sqrt{2} - 8\right)}{48} & - \frac{2\, \sqrt{3}}{3} & - \frac{\sqrt{3}\, \left(9\, \sqrt{2} + 8\right)}{48} & \frac{\sqrt{3}\, \left(9\, \sqrt{2} - 8\right)}{48} & - \frac{67\, \sqrt{3}}{96} & \frac{\sqrt{3}\, \left(9\, \sqrt{2} - 8\right)}{48} & - \frac{2\, \sqrt{3}}{3} & - \frac{\sqrt{3}\, \left(9\, \sqrt{2} + 8\right)}{48} & - \frac{67\, \sqrt{3}}{96}\\ 1 & -1 & 0 & 0 & 0 & 0 & \frac{7\, \sqrt{2}}{16} & - \frac{7\, \sqrt{2}}{16} & 0 & \frac{7\, \sqrt{2}}{16} & - \frac{7\, \sqrt{2}}{16} & \frac{13}{32} & - \frac{7\, \sqrt{2}}{16} & 0 & \frac{7\, \sqrt{2}}{16} & \frac{13}{32}\\ 0 & 0 & \frac{1}{4} & - \frac{1}{4} & 0 & 0 & - \frac{\sqrt{2}}{16} & 0 & - \frac{\sqrt{2}}{8} & \frac{\sqrt{2}}{16} & \frac{\sqrt{2}}{16} & - \frac{5}{32} & 0 & - \frac{\sqrt{2}}{8} & 0 & \frac{5}{32}\\ 0 & 0 & \frac{3}{4} & \frac{3}{4} & - \frac{3}{4} & - \frac{3}{4} & \frac{3}{8} - \frac{3\, \sqrt{2}}{16} &  - \frac{3\, \sqrt{2}}{16} - \frac{3}{8} & 0 & \frac{3}{8} - \frac{3\, \sqrt{2}}{16} & \frac{3\, \sqrt{2}}{16} + \frac{3}{8} & - \frac{3}{32} &  - \frac{3\, \sqrt{2}}{16} - \frac{3}{8} & 0 & \frac{3\, \sqrt{2}}{16} - \frac{3}{8} & - \frac{3}{32}\\ 0 & 0 & \frac{9}{4} & - \frac{9}{4} & 0 & 0 & - \frac{9\, \sqrt{2}}{16} & 0 & \frac{9\, \sqrt{2}}{8} & \frac{9\, \sqrt{2}}{16} & \frac{9\, \sqrt{2}}{16} & \frac{27}{32} & 0 & \frac{9\, \sqrt{2}}{8} & 0 & - \frac{27}{32}\\ 0 & 0 & \frac{3}{4} & - \frac{3}{4} & 0 & 0 & \frac{3}{4} - \frac{9\, \sqrt{2}}{16} & 0 & \frac{3\, \sqrt{2}}{8} & \frac{9\, \sqrt{2}}{16} - \frac{3}{4} & \frac{9\, \sqrt{2}}{16} + \frac{3}{4} & \frac{3}{32} & 0 & \frac{3\, \sqrt{2}}{8} & 0 & - \frac{3}{32}\\ 0 & 0 & \frac{3}{4} & - \frac{3}{4} & 0 & 0 &  - \frac{9\, \sqrt{2}}{16} - \frac{3}{4} & 0 & \frac{3\, \sqrt{2}}{8} & \frac{9\, \sqrt{2}}{16} + \frac{3}{4} & \frac{9\, \sqrt{2}}{16} - \frac{3}{4} & \frac{27}{32} & 0 & \frac{3\, \sqrt{2}}{8} & 0 & - \frac{27}{32}\\ 0 & 0 & \frac{3}{4} & \frac{3}{4} & - \frac{3}{4} & - \frac{3}{4} & \frac{3\, \sqrt{2}}{16} + \frac{3}{8} & \frac{3\, \sqrt{2}}{16} - \frac{3}{8} & 0 & \frac{3\, \sqrt{2}}{16} + \frac{3}{8} & \frac{3}{8} - \frac{3\, \sqrt{2}}{16} & - \frac{9}{32} & \frac{3\, \sqrt{2}}{16} - \frac{3}{8} & 0 &  - \frac{3\, \sqrt{2}}{16} - \frac{3}{8} & - \frac{9}{32}\\ 0 & 0 & 0 & 0 & - \frac{1}{4} & \frac{1}{4} & 0 & - \frac{\sqrt{2}}{16} & - \frac{\sqrt{2}}{8} & 0 & 0 & \frac{\sqrt{2}}{32} & \frac{\sqrt{2}}{16} & \frac{\sqrt{2}}{8} & - \frac{\sqrt{2}}{16} & \frac{\sqrt{2}}{32}\\ 0 & 0 & 0 & 0 & 0 & 0 & 0 & 0 & \frac{3}{4} & 0 & 0 & - \frac{3\, \sqrt{2}}{16} & 0 & - \frac{3}{4} & 0 & \frac{3\, \sqrt{2}}{16}\\ 0 & 0 & 0 & 0 & \frac{9}{4} & - \frac{9}{4} & 0 & \frac{9\, \sqrt{2}}{16} & - \frac{9\, \sqrt{2}}{8} & 0 & 0 & \frac{27\, \sqrt{2}}{32} & - \frac{9\, \sqrt{2}}{16} & \frac{9\, \sqrt{2}}{8} & \frac{9\, \sqrt{2}}{16} & \frac{27\, \sqrt{2}}{32}\\ 0 & 0 & 0 & 0 & - \frac{3}{4} & \frac{3}{4} & 0 &  - \frac{9\, \sqrt{2}}{16} - \frac{3}{4} & \frac{3\, \sqrt{2}}{8} & 0 & 0 & - \frac{3\, \sqrt{2}}{32} & \frac{9\, \sqrt{2}}{16} + \frac{3}{4} & - \frac{3\, \sqrt{2}}{8} & \frac{3}{4} - \frac{9\, \sqrt{2}}{16} & - \frac{3\, \sqrt{2}}{32}\\ 0 & 0 & 0 & 0 & \frac{3}{4} & - \frac{3}{4} & 0 & \frac{9\, \sqrt{2}}{16} - \frac{3}{4} & - \frac{3\, \sqrt{2}}{8} & 0 & 0 & \frac{27\, \sqrt{2}}{32} & \frac{3}{4} - \frac{9\, \sqrt{2}}{16} & \frac{3\, \sqrt{2}}{8} & \frac{9\, \sqrt{2}}{16} + \frac{3}{4} & \frac{27\, \sqrt{2}}{32}\\ 0 & 0 & 0 & 0 & 0 & 0 & 0 & 0 & - \frac{3}{4} & 0 & 0 & \frac{9\, \sqrt{2}}{16} & 0 & \frac{3}{4} & 0 & - \frac{9\, \sqrt{2}}{16} 
	\end{array}
	\right)
	\end{array}
\end{equation}

%\end{landscape}
%\global\pdfpageattr\expandafter{\the\pdfpageattr/Rotate 90}
\end{sidewaysfigure}
\clearpage
\twocolumngrid
\begin{acknowledgments}
	The authors would like to thank Professor Daniel James from the University of Toronto for useful discussions. This work was supported by grants from the Natural Sciences and Engineering Research Council of Canada and the Canadian Institutes of Health Research. M.S. acknowledges the financial support from the Ontario Graduate Scholarship.
\end{acknowledgments}
\bibliography{THGPRBIB}
\end{document}